%% file: main.tex
\documentclass[aps,onecolumn,superscriptaddress,nofootinbib,a4paper,longbibliography,pra,10pt]{revtex4-2}

\usepackage{graphicx}
\usepackage{dcolumn}
\usepackage{bm}
\usepackage{multirow}%
\usepackage{amsmath}
\usepackage{amssymb}
\usepackage{amsfonts}%
\usepackage{amsthm}%
\usepackage{mathrsfs}%
\usepackage{xcolor}%
\usepackage{booktabs}%
\usepackage{listings}%
\usepackage{physics}
\usepackage{dsfont}
\usepackage{soul}
\usepackage[utf8]{inputenc}
\usepackage{braket}
\usepackage{mathtools}
\usepackage[normalem]{ulem}

\newcommand{\RomanNumeralCaps}[1]
    {\MakeUppercase{\romannumeral #1}}

\usepackage{subcaption}

\usepackage{pifont} 

\newcommand{\beginsupplement}{%
        \setcounter{table}{0}
        \renewcommand{\thetable}{S\arabic{table}}%
        \setcounter{figure}{0}
        \renewcommand{\thefigure}{S\arabic{figure}}%
        \renewcommand{\theequation}{S.\arabic{equation}}
     }

\newcommand{\cmark}{\ding{51}}%
\newcommand{\xmark}{\ding{55}}%

\usepackage[colorlinks=true,linkcolor=blue,citecolor=magenta,urlcolor=blue]{hyperref}

\begin{document}



\title{
Imperceptible measurement deviations can fake high-dimensional entanglement}

\author{Jaime Moreno}
\email{Contact author: jaime.morenozuleta@tuni.fi}
\affiliation{Photonics Laboratory, Physics Unit, Tampere University, Tampere, Finland}

\author{Elna Svegborn}
\affiliation{Physics Department and NanoLund, Lund University, Lund, Sweden}

\author{Simon Morelli}
\affiliation{Atominstitut \& Vienna Center for Quantum Science and Technology (VCQ), TU Wien, Vienna, Austria}

\author{Markus Hiekkamäki}
\affiliation {Quantum Technology Laboratories GmbH, Vienna, Austria}

\author{Lea Kopf}
\affiliation{VTT Technical Research Centre of Finland, Espoo, Finland}

\author{Robert Fickler}
\affiliation{Photonics Laboratory, Physics Unit, Tampere University, Tampere, Finland}

\author{Armin Tavakoli}
\email{Contact author: armin.tavakoli@fysik.lu.se}
\affiliation{Physics Department and NanoLund, Lund University, Lund, Sweden}

\date{\today}

\begin{abstract}
Quantum entanglement is a central resource for emerging quantum technologies, but its experimental certification usually assumes that prescribed quantum measurements can be realized exactly. This assumption becomes increasingly restrictive in high-dimensional systems, where accurate measurement control is more challenging. Here, we show that arbitrarily small, adversarially tailored measurement deviations can lead to the false certification of high-dimensional entanglement in systems that are, in fact, separable, with the effect becoming increasingly severe as the dimension grows. We experimentally demonstrate this across a broad range of dimensions using classical photonic states encoded in the spatial degree of freedom. We report false certification of several tens of entangled dimensions in systems spanning up to 61  dimensions while exploiting measurement deviations as low as $0.23\%$. Our results reveal how imperceptible invasive attacks can critically compromise high-dimensional entanglement certification, motivating the development of new certification methods robust to bounded discrepancies between theory and experiment.

\end{abstract}


\maketitle

\textit{Introduction.---} Conventional entanglement certification assumes that prescribed quantum observables can be implemented with perfect accuracy \cite{G_hne_2009}. In realistic experiments, however, measurement implementations inevitably deviate from their idealized descriptions, with accurate control becoming increasingly challenging as the system dimension grows \cite{Friis2019, Erhard2020, valencia2020high}. Consequently, certification methods that assume ideal high-dimensional measurements become increasingly susceptible to misleading conclusions. Thus, possible experimental deviations, unless they are properly accounted for, have particularly adverse impact on high-dimensional entanglement verification, which underlies emerging capabilities in quantum communication \cite{Cozzolino2019, chang2026highdimensionalquantumcommunicationscalable}, quantum computation \cite{Reimer2019, Ringbauer2022}, and quantum networks \cite{zheng2023multichip, valencia2025large}. Several avenues in quantum information address measurements that feature noise, systematic biases, or adversarial attacks \cite{Brunner2014, Brask2026, Oliveira2026, Uola2020}. To remove such concerns, device-independent approaches treats measurement devices as uncharacterized, but this comes at the cost of reduced certification power and more demanding experimental requirements, particularly in high dimensions. 

A practical middle ground is to quantify the experimenter's degree of measurement control while allowing the remaining uncertainty to be uncharacterized, including adversarial attacks. Existing approaches along these lines have focused largely on qubit systems \cite{Rosset_2012, Cao_2024}, while beyond qubits available methods rely primarily on numerical analysis at low dimensions and offer little insight into the scaling of entanglement certification \cite{Morelli_2022, DAlessandro2026}. This leaves an important question unanswered: can measurement deviations that lie within the residual uncertainty of even state-of-the-art high-dimensional experiments nevertheless lead to sizable false certifications of entanglement?

\begin{figure*}[ht]
    \centering
    \includegraphics[width=0.8\textwidth]{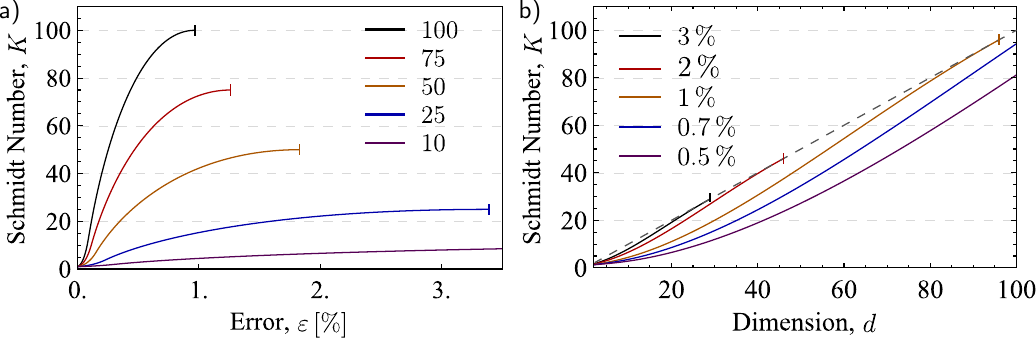}
    \caption{
        Falsely certified Schmidt number $K$ as a function of:
        (a) error for fixed dimension $d\in\{10,25,50,75,100\}$, and
        (b) dimension for fixed error
        $\varepsilon\in\{0.5\%,0.7\%,1\%,2\%,3\%\}$.
        The gray dashed line in (b) corresponds to the largest meaningful
        Schmidt number ($K=d$) under the idealized measurement assumptions.
        The cut-off at which the model reaches $K=d$ is marked in both
        panels by vertical lines.}
    \label{fig:average_protocol}
\end{figure*}

Here, we show that the answer indeed is affirmative,  by considering one of the most widely used witness tests of high-dimensional entanglement \cite{Spengler_2012}. We show that quantum measurements that  remain arbitrarily close to their target observables still can produce increasingly large false certifications of entanglement as the dimension grows, even when the measurements are performed on states that are entirely classical.
This vulnerability persists across three distinct and experimentally motivated models of measurement deviations, demonstrating that the effect is not tied to a particular measure of experimental error. We experimentally demonstrate the practical relevance of this effect using product states encoded in the spatial modes of attenuated laser pulses, falsely certifying substantial entanglement dimensionality over a broad range of system dimensions and measurement deviations. For example, we falsely certify 26-dimensional entanglement using only a measurement deviation of  $0.7\%$ in a 61-dimensional classical state. Our work illustrates how current standard methods for high-dimensional entanglement certification are vulnerable to tiny measurement deviations and motivate the need for appropriate certification frameworks and tests for taking such deviations into account.


\textit{Entanglement witness and $\varepsilon$-invasive attacks.---}\label{sec2}
Consider that Alice and Bob share a bipartite state $\rho_{AB}$ with local dimension $d$. Their goal is to certify its entanglement dimensionality, quantified by the Schmidt number. A state has Schmidt number $K$ if its correlations require entanglement across exactly $K$ local levels \cite{Terhal_2000}.

A standard approach to lower-bound the Schmidt number is through entanglement witnesses \cite{Sanpera2001, Chruciski2014}. In this case, Alice and Bob perform prescribed local measurements on their share of the state $\rho_{AB}$ and then use the resulting correlations to infer a lower bound on $K$. Witnesses that are experimentally practical in arbitrary dimension are scarce, but a prominent example requires measurements in only two mutually unbiased bases \cite{Spengler_2012, Bavaresco2018}. The simplest choice of such bases is the computational basis $\{\ket{l}\}_{l=0}^{d-1}$ and its Fourier transform $\{\ket{f_l}=G\ket{l}\}_{l=0}^{d-1}$, where $G=\frac{1}{\sqrt d}\sum_{m,n=0}^{d-1}e^{\frac{2\pi i}{d}mn}\ketbra{m}{n}$. The witness sums the probabilities of correlated outcomes in the two bases \cite{Spengler_2012, Morelli_2023},
\begin{align}\label{entanglementwitness}
W_d ={}& \sum_{l=0}^{d-1} \Big[
\Tr\left(\ketbra{l}{l} \otimes \ketbra{l}{l}\rho_{AB}\right)
\nonumber\\
&\quad +
\Tr\left(\ketbra{f_l}{f_l} \otimes
\ketbra{f_l^*}{f_l^*}\rho_{AB}\right)
\Big]
\leq 1+\frac{K}{d},
\end{align}
where $\ket{f_l^*}$ is the complex conjugate of $\ket{f_l}$. The bound holds for all states with Schmidt number at most $K$ and a violation certifies a Schmidt number of at least $K+1$.

Crucially, Eq~\eqref{entanglementwitness} assumes that the measurements are implemented exactly. In an experiment, however, Alice and Bob instead realize imperfect approximations of the prescribed measurements. To quantify the agreement between an arbitrary target basis $\mathbf{O}=\{\ketbra{O_l}{O_l}\}_{l=0}^{d-1}$ and  its associated laboratory implementation $\tilde{\mathbf{O}}=\{\tilde{O}_l\}_{l=0}^{d-1}$ we use the average measurement fidelity
\begin{equation}\label{fidelityconstraint}
\mathcal{F}(\mathbf{O},\tilde{\mathbf{O}})
\equiv \frac{1}{d}\sum_{l=0}^{d-1}
\langle O_l|\tilde{O}_l|O_l\rangle
\geq 1-\varepsilon .
\end{equation}
Here, the imprecision parameter $\varepsilon \geq 0$ bounds the experimentally unresolved part of the measurement implementation, i.e.,~the device is guaranteed to agree with its target up to fidelity $1-\varepsilon$. The remaining deviation is treated in a device-independent manner, without any assumptions about its origin, whether it arises from arbitrary noise or adversarial influence. We refer to exploiting this unresolved freedom while remaining within the experimentally tolerated deviation as an $\varepsilon$-invasive attack. The purpose of such an attack is to induce false violations of the high-dimensional entanglement witness, while operating below the level of resolution available to the experimenter. The question we address is whether or not an $\varepsilon$-invasive attack can cause far greater errors in the conclusions of the test than it introduces in the measurements.

\textit{Faking high Schmidt number.---} We construct an explicit $\varepsilon$-invasive attack for arbitrary dimension $d$ that is tailored to the witness $W_d$ in Eq~\eqref{entanglementwitness}. The computational basis is deviated into an $\varepsilon$-close basis $\{\ket{\phi_k}\}_k$ of the form
\begin{equation}\label{perturbedbasis}
    \begin{aligned}
   &\ket{\phi_{0}} = \alpha \ket{0}+ \beta \sum_{j=1}^{d-1}\ket{j}, \\
   &\ket{\phi_k} = \beta \ket{0} - \gamma \ket{k} + \delta  \sum_{\substack{j = 1\\j \neq k}}^{d-1} \ket{j},
    \end{aligned}
\end{equation}
for $k=1,\ldots,d-1$. Similarly, the Fourier basis is deviated as $\{G\ket{\phi_k}\}_k$. In Supplementary Material (SM), we derive coefficients  to be 
\begin{equation}
\begin{aligned}
        &\alpha = \frac{2(d-1)R-(d-2)}{d}, \quad  &\gamma = \frac{2R+d-2}{d}  \\
        &\beta = \frac{2}{d}\sqrt{(1-R)(1+(d-1)R)}, &\quad   \delta = \frac{2}{d}(1-R).
\end{aligned}
\end{equation}
where $ R(\varepsilon,d) \equiv (1-\frac{\varepsilon d^2}{4(d-1)})^{1/2}$. This construction is valid in the regime $0\leq \varepsilon\leq 4(d-1)/d^2$ where $R\geq 0$. Although decreasing, the upper threshold of $\varepsilon$ remains relevant up to large $d$ in the small-$\varepsilon$ regime that we are interested in. Since we aim to fake the certification of entanglement, we apply this attack to a product state, $\ket{\psi}_A\otimes\ket{\psi}_B$. The witness then becomes $\mathcal{W}^{(d)}_{\text{prod}}(\varepsilon)=  \sum_{k=0}^{d-1} \left(\lvert\braket{\psi|\phi_k}\rvert^4 + \lvert\braket{\psi|G|\phi_k}\rvert^4 \right)$.
To maximize this, we set $\ket{\psi} =  \sqrt{1-(d-1)\mu^2} \ket{0}+\mu \sum_{j=1}^{d-1}\ket{j}$. The optimal choice of $\mu$ is obtained by differentiating $\mathcal{W}^{(d)}_{\text{prod}}$ w.r.t $\mu$ and then for a fixed $d$ and $\varepsilon$ numerically finding the appropriate root of the derivative.

Despite the measurements remaining $\varepsilon$-close to their targets, this attack produces witness values that falsely certify remarkably high Schmidt numbers. The results are illustrated in Fig~\ref{fig:average_protocol}, where we showcase the faking capability for small values of $\varepsilon$. We see that the effect becomes rapidly more pronounced as the dimension increases. For example, in dimension $d=61$, a deviation of only $\varepsilon=0.7\%$ is sufficient for a product state to falsely certify a Schmidt number of at least $47$. Most strikingly, for any fixed $\varepsilon>0$, a sufficiently large dimension allows a product state to attain the same witness value as a maximally entangled state under ideal measurements. Thus, arbitrarily small measurement deviations can ultimately fake even maximal entanglement.


\begin{figure}
    \centering
    \includegraphics[width=0.96\linewidth]{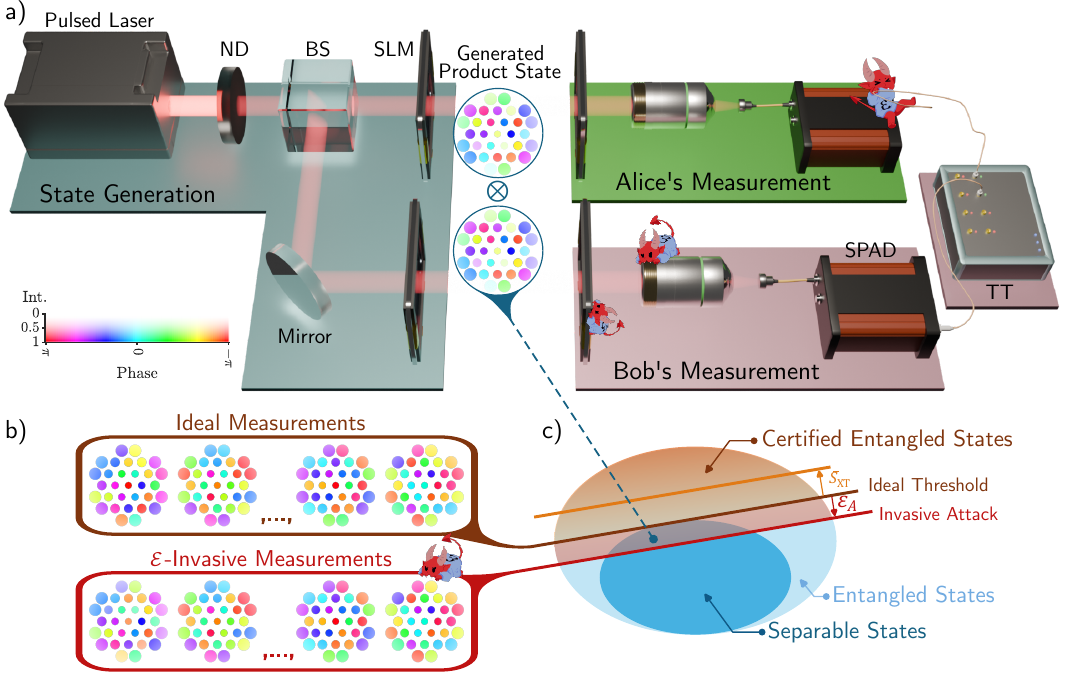}
    \caption{Illustrative schematic of the experimental setup employing the macro-pixel basis as the local degree of freedom. a) Simplified experimental setup. An attenuated pulsed laser and a BS serves as a source of uncorrelated photon pairs, and two SLMs are used to prepare a $d$-dimensional separable state in the macro-pixel basis. Alice and Bob’s measurements are carried out using another pair of SLMs followed by SMFs. b) Computational basis states of the ideal (top) and $\varepsilon$-invasive (bottom) measurement bases, shown for $\varepsilon = 0.7 \%$. The $\varepsilon$-demons illustrate the unresolved deviations introduced by an attacker. c) Classification of quantum states into entanglement classes by an entanglement witness under $\varepsilon$-invasive attack and average per-mode crosstalk $\bar{S}_{XT}$.    ND: Neutral density filter, BS: Beam splitter, SLM: Spatial light modulator, SPAD: Single-photon avalanche diode, SMF: Single mode fiber, TT: Time tagger.}
    \label{fig:mainsetup}
\end{figure}

\textit{Experiment.---} To demonstrate this phenomenon, we conduct a series of experiments in which separable states, subjected to $\varepsilon$-invasive measurement deviations far smaller than the imperfections present in current high-dimensional quantum platforms, nevertheless produce false certifications of large amounts of high-dimensional entanglement.

To guarantee that the experimentally generated photonic states are fully uncorrelated, we use an attenuated pulsed laser. The laser operates at a central wavelength of 780\,nm with a pulse duration of 2.5\,ps.  After attenuation, the laser pulse is split into two distinct spatial paths, as shown in Fig~\ref{fig:mainsetup}\,a). The short pulse duration allows us to correlate detection events in time and thereby reproducing the coincidence detection scheme used in experiments with true photon pairs. The desired product state $\ket{\psi}_A\otimes\ket{\psi}_B$ is prepared by independently applying holographic phase and amplitude modulation  to the light in each arm using spatial light modulators (SLMs) \cite{Bolduc:13}.

We encode the high-dimensional quantum states in the transverse spatial degree of freedom (DOF) of light. This continuous DOF is mapped to a finite-dimensional space using macro-pixels as basis states \cite{valencia2020high}, with each macro-pixel (disk) a distinct quantum state. Arbitrary superpositions are generated by controlling the amplitudes and phases of each macro-pixel as illustrated by the optimal separable state in Fig~\ref{fig:mainsetup}\,a)  associated to the $\varepsilon$-invasive attack in Eq~(\ref{perturbedbasis}) for dimension 31. 

Alice and Bob's measurements are implemented through holographic modulation in a SLM followed by projection into a single-mode fiber (SMF)~\cite{Bouchard_2018}. 
The $\varepsilon$-demons represent small perturbations beyond Bob and Alice's control that exploit the unresolved discrepancies between the actual measurement and their idealized description. Fig~\ref{fig:mainsetup}\,b) showcases several of the computational basis projectors constructed using Eq~(\ref{eqEM:computationalbasis_change}) in dimension 31, together with their $\varepsilon$-invasive counterparts for $\varepsilon =0.7\%$. 
The small deviations between the two measurements are visible by phase or amplitude differences of the  states. The perturbation is predominantly localized on the first state due to the construction in Eq~(\ref{perturbedbasis}), whereas the remaining projectors are mostly unperturbed. We note, however, that this asymmetry can be overcome by leveraging classical randomness between Alice's and Bob's devices (see SM for details). 

The joint measurements $\ketbra{\phi_k}{\phi_k}_A\otimes\ketbra{\phi_l}{\phi_l}_B$, are implemented through individual spatial mode projections. Ideally, the high-dimensional quantum measurements in Eq~\eqref{entanglementwitness} require multi‑outcome projectors $\{\ketbra{\phi_k}{\phi_k}\}_k$ \cite{Tavakoli_2025}. 
However, realizing a high-dimensional multi-outcome measurement with sufficient fidelity and control to implement the proposed $\varepsilon$-deviations remain challenging in current photonic architectures. Instead, we implement the measurement through a sequence of individual local projectors $\ketbra{\phi_k}{\phi_k}$ and reconstruct the corresponding outcome probabilities from the count rates of the single-photon avalanche diodes (SPADs) and time tagger.
The measurements in the Fourier basis only require modifying the displayed holograms on Alice’s and Bob’s SLMs.
Further implementation details can be found in the SM.


\begin{figure*}
    \centering
    \includegraphics[width=0.8\textwidth]{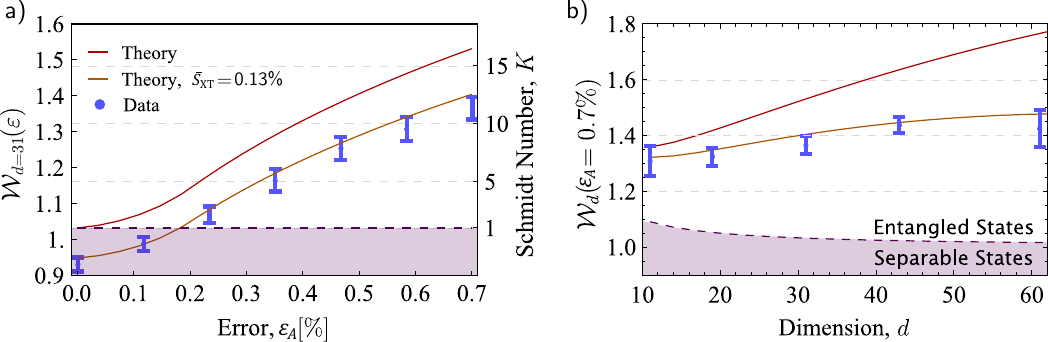}
    \caption{
    Experimental results. a) scaling of the entanglement witness with $\varepsilon_A$ at a fixed local dimension $d = 31$. The red curve indicates the ideal theoretical model, whereas the orange curve shows the model after including a average per-mode crosstalk of $\bar{S}_{XT} = 0.13\%$. b) Scaling of the entanglement witness value with dimension for a fixed perturbation $\varepsilon_A=0.7\%$.}
    \label{fig:Exp_aveprotocol}
\end{figure*}




Intrinsic experimental imperfections, including dark counts, misalignment, optical aberrations, and imperfect SLM–SMF mode filtering, produce small non-zero overlap $S_{i\neq j} = \left|\braket{i|j}\right|^2$, known as crosstalk. Consequently, the average fidelity between the target and implemented measurement bases, defined in Eq~(\ref{fidelityconstraint}), is less than one already before the attack $\varepsilon$-invasive is introduced. That is, the intrinsic imperfections introduce an additional deviation on top of the $\varepsilon$-invasive attack. As a result, $\mathcal{F}(\mathbf{O},\tilde{\mathbf{O}}) = \bar{S}_D (1-\varepsilon_A) = (1-\varepsilon)$, where $\bar {S}_D$ is the average detection probability defined in the End Matter,  $\varepsilon_A$ is the systematic deviation introduced by the attacker, and $\varepsilon$ is the total deviations from the target measurement.

We characterize the intrinsic imprecisions in the experiment using the average per-mode crosstalk $\bar{S}_{XT}$. Since $\bar{S}_{D} = 1- (d-1)\bar{S}_{XT}$, the same per-mode crosstalk produces a larger total infidelity in higher-dimensional systems.
We measure an average per-mode crosstalk of $\bar{S}_{XT} = 0.12 \pm 0.01\%$ for $d=11$, 19, 31, 43, and 61. Theoretical predictions assuming a $\bar{S}_{XT} = 0.13\%$ fit the experimental data the best.
We attribute this small discrepancy to mechanical and thermal drifts that gradually increase the crosstalk and degrade the measurement fidelity during the measurement period.

From the experimenter's perspective, only the total infidelity relative to the target measurement is directly accessible. Ordinary imperfections, such as noise and crosstalk, reduce the effectiveness of the witness but do not themselves produce false positives. By contrast, an $\varepsilon$-invasive attack shifts the effective witness boundary into the set of separable states, as illustrated in Fig~\ref{fig:mainsetup}\,c), while remaining small enough to be concealed by the natural measurement imperfections.

Fig~\ref{fig:Exp_aveprotocol}\,a) presents the results for $d = 31$ and attack strengths up to $\varepsilon_A = 0.7\%$. False certifications of entanglement occur for perturbations as small as $\varepsilon_A = 0.23\%$. 
At $\varepsilon_A = 0.7\%$ we measure $\mathcal{W}_d = 1.36 \pm 0.03 $, corresponding to a falsely certified Schmidt number of $K = 11$. Although, the measured witness values are below the theoretical predictions, the results agree well with theory when experimental crosstalk is included.

The theory predicts that $\varepsilon$-invasive attacks become more powerful with growing dimension. We test this dependence by fixing $\varepsilon_A = 0.7\%$ and varying the dimensionality of our system up to $d = 61$, as shown in Fig~\ref{fig:Exp_aveprotocol}\,b).
In all measured dimensions, we are able to exceed the separable-state threshold. However, experimental crosstalk decreases the highest measurable $K$, such that the discrepancy between the measured witness value and the ideal theoretical prediction increases for higher dimensions. Nevertheless, we are able to falsely certify a Schmidt number as large as $K = 26$ for $d=61$. 

From the measured per-mode crosstalk, the total measurement infidelity in the absence of an attack is $3.6\pm0.1\%$ for $d=31$ and $8.5\pm0.2\%$ for $d=61$; additional details are given in the SM. Introducing an attack with $\varepsilon_A=0.7\%$ increases these values to $\varepsilon=4.3\%$ and $\varepsilon=9.1\%$ for dimension 31 and 61, respectively.
Thus, the attack produces only a modest change in the experimentally accessible measurement infidelity while still yielding a large false certification of entanglement. More strikingly, at $d=31$ an attack as small as $\varepsilon_A=0.23\%$, corresponding to only an approximately $6\%$ relative increase from the intrinsic infidelity, is sufficient to violate the separable-state bound.

\textit{Other quantifiers of measurement fidelity.---} We have focused on quantifying measurement deviations through the average fidelity \eqref{fidelityconstraint}, but depending on how a measurement is implemented other fidelity quantifiers may become relevant. For instance, one may consider the fidelity at the level of each measurement operator, namely $\langle O_l|\tilde{O}_l|O_l\rangle \geq 1-\varepsilon$ $\forall l$. This is more restrictive than our model in Eq~\eqref{fidelityconstraint}.  Another option is to consider the fidelity error allowed for each click-event \cite{Tavakoli_2025}, relevant when each outcome of the target basis is measured through a distinct individual projection, i.e., when $\{O_l\}$ is binarized into $d$ separate click or no-click measurements. Naturally, changing the quantifier of the measurement deviation must affect the design and power of the best $\varepsilon$-invasive attacks. In the SM, we discuss these models in detail and derive explicit $\varepsilon$-invasive attacks for them (section \RomanNumeralCaps{2}.B and \RomanNumeralCaps{2}.C). We have also extended our experimental tests to these attacks (section \RomanNumeralCaps{4}.F). While the specific amount of false entanglement varies, the conclusions are qualitatively similar to those we have showcased here for the the average fidelity quantifier given by Eq~\eqref{fidelityconstraint}. Our analysis for the three distinct quantifiers suggests  that the vulnerability of high-dimensional entanglement certification does not arise from a particular experimental implementation or fidelity choice, but it is intrinsic to the conventional paradigm for entanglement witnessing. Importantly, in two of the models the $\varepsilon$-invasive attacks can exploit classical randomness to reproduce not only the value of the witness \eqref{entanglementwitness}, but also the full probability distribution expected from an honest implementation with entanglement. Thus, no alternative witness based on the same measurement statistics can circumvent these attacks.

Moreover, evidence indicates that the families of $\varepsilon$-invasive attacks that we have designed for all three measurement error models are essentially the best ones possible. This is based on systematic numerical search up to twenty dimensions and random choices of $\varepsilon$ (see SM).

\textit{Conclusions.---}
Our results reveal a significant vulnerability of conventional high-dimensional entanglement certification to minimal measurement deviations. Here, we show that measurement deviations below the resolution of state-of-the-art quantum technology can be exploited adversarially to produce outcome statistics that falsely pass conventional tests of high-dimensional entanglement. The magnitude of this effect can be striking, with minute deviations leading to the false certification of large numbers of entangled dimensions and, at sufficiently large dimension, even reproducing the statistics associated with maximal entanglement using mere product states. Importantly, the vulnerability grows with dimension, while experimental control typically deteriorates.

The implications may extend to QKD, as certification methods based on mutually unbiased bases are closely related to protocols such as BB84, given that single photon prepare-and-measure schemes can often be mapped to entanglement-based ones \cite{Cerf2002}. We anticipate that minimal but carefully designed perturbations in the preparation and measurement apparatuses can significantly undermine the required security thresholds on both forms of QKD.

More broadly, Our findings give further motivation for device-independent approaches to quantum information \cite{zapatero2023advances}. They show that even when a measurement is almost fully controlled, the small unresolved remainder can suffice to provide an adversary with significant power. This raises a natural question for high-dimensional quantum information. Can entanglement witnesses be designed to remain secure against $\varepsilon$-invasive attacks without substantially sacrificing their certification power? Our results suggest that achieving both may primarily require physically motivated restrictions on the unresolved measurement deviations, or potentially  substantially different witness methods. Developing such robust methods is relevant for reliably exploiting the growing scale of emerging high-dimensional quantum technologies.

\begin{acknowledgments}
We thank Valerio Scarani for useful discussions. E.S and A.T are supported by the Knut and Alice Wallenberg Foundation through the Wallenberg Center for Quantum Technology (WACQT), the Swedish Research Council under Contract No. 2023-03498 and the Swedish Foundation for Strategic Research.
J.M and R.F acknowledge the financial support of the Finnish Ministry of Education and Culture through the PREIN/I-DEEP doctoral pilot (VN/3137/2024-OKM-4). We also acknowledge the Flagship for Photonics Research and Innovation (PREIN), decision number Research Council of Finland/2024/368650.
S.M is a recipient of an APART-MINT Fellowship of the Austrian Academy of Sciences at the Atominstitut of the Technische Universit\"at Wien.
L.K acknowledges support by the Postdoctoral Programme for Research Institutes in Finland funded by the Finnish Government.
\end{acknowledgments}

\section*{End Matter}\label{secEM}
\textit{Definition of the computational basis.---}\label{sec11}
The natural computational basis of the macro-pixel states consist by single disks, where as the  associated Fourier basis would then be constructed as superpositions of $d$ disks (see section~4.1 in the SM). 
The states in this computational basis would have significantly lower intensities than the states in the Fourier basis.
These large intensity differences would make the experiment unfeasible.
To circumvent this issue we introduce a basis transformation $U_x$ since the entanglement witness in Eq~\eqref{entanglementwitness} does not change upon local unitary transformations of both input state and measurement bases. The new computational basis is given by
\begin{equation}\label{eqEM:computationalbasis_change}
    U_x = \frac{1}{\sqrt{d}}\sum_{k=0}^{d-1} X^k\omega^{\frac{k}{2}(k+1)},
\end{equation}
with $\omega = \exp(2\pi i/d)$ and the $d$-dimensional $X$-gate. The transformed states, shown in  Fig~\ref{fig:mainsetup}\,a) and c), are such that both computational and Fourier states are superpositions of $d$ disks equalizing the intensity.

\textit{Fidelity constraints including crosstalk.---}
For a set of projectors $\{O_0,O_1, ...., O_{d-1}\}$ that define a quantum measurement, crosstalk acts as a convex mixing of the elements in the set. Therefore, the implemented quantum measurement can be described as
\begin{equation}\label{eq:crosstalkE}
    \tilde{O}_i = \sum_{j=0}^{d-1} S_{ij}O_j,
\end{equation}
where $S_{ij} = \text{Tr}(O_j^\dagger\tilde{O}_i)$ are the elements of the crosstalk matrix $S$. With this relation, one can compute the new bounds for the $\varepsilon$-deviation for the models containing the experimental crosstalk as
\begin{equation}\label{eq_end:ct_avemodel}
    \frac{1}{d}\sum_{j=0}^{d-1} \langle j|\tilde{O}_j|j\rangle = \braket{S}(1- \varepsilon),  
\end{equation}
where we replaced the individual terms $S_{jj}$ with the average detection probability $\bar{S}_D = \frac{\text{Tr}(S)}{d}$.  The average per-mode crosstalk is defined as $\bar{S}_{XT} = \frac{1}{d(d-1)}\sum\limits_{i,j = 0, i\neq,j}^{d-1}S_{ij} $.




\newpage
\input{supplementary}

\bibliography{references}
\end{document}

%% file: supplementary.tex
\beginsupplement{

\section{Entanglement detection}

\subsection{Schmidt numbers}
Consider a bipartite state $\rho_{AB}$, with subsystems of local dimension $d_A$ and $d_B$, respectively. The composite system is called separable if and only if it can be written as a convex combination of product states
\begin{equation}\label{eq:sep}
    \rho_{AB} = \sum_\lambda  q_\lambda \ \phi_\lambda \otimes \varphi_\lambda,
\end{equation}
for some probability distribution $\{q_\lambda\}_\lambda$ and arbitrary states $\phi_\lambda$ and $\varphi_\lambda$ on subsystem $A$ and $B$, respectively \cite{G_hne_2009}. If $\rho_{AB}$ cannot be decomposed in the form of Eq.~\eqref{eq:sep}, it is called entangled. However, knowing that a pair of high-dimensional systems are entangled does not reveal whether all internal levels of the two systems are involved in generating the entanglement. The common way to address this is to analyze not only whether entanglement is present but also the dimensionality of the entanglement.

For bipartite states, the entanglement dimensionality is represented by the Schmidt number $K$. Formally, the Schmidt number of a density matrix $\rho_{AB} = \sum_\lambda p_\lambda \ketbra{\psi_\lambda}{\psi_\lambda}$ is defined as the largest Schmidt rank of all pure states $\ket{\psi_\lambda}$, minimized over all possible ensemble-decompositions of the density matrix, where $\{p_\lambda \geq0, \ket{\psi_\lambda}\}$ \cite{Terhal_2000}. For a pure state $\ket{\psi}_{AB}$, the Schmidt rank $r(\psi)$ is the rank of its marginals, or equivalently the number of terms appearing in its Schmidt decomposition, $\ket{\psi} = \sum_{i=1}^{r(\psi)} \lambda_i \ket{\alpha_i,\beta_i}$, where $\{\ket{\alpha_i}\}_i$ and $\{\ket{\beta_i}\}_i$ are orthonormal sets of states, and the coefficients satisfy $\lambda_i > 0$ and $\sum_i \lambda_i^2 = 1$. Thus, the Schmidt number of $\rho_{AB}$ is defined as
\begin{equation}
    K(\rho_{AB}) = \min_{\{p_\lambda,\psi_\lambda\}} \quad \left\{r_{\max}: \rho_{AB} = \sum_\lambda p_\lambda \ketbra{\psi_\lambda}{\psi_\lambda} \quad \text{and} \quad r_{\max} = \max_\lambda r(\psi_\lambda) \right\}.
\end{equation}
The Schmidt number is an integer in the range $1 \leq K \leq \min\{d_A,d_B\}$. Note, that the case of $K = 1$ is equivalent to separability, whereas all states with $K>1$ are entangled. The specific value of $K$ is the entanglement dimension of $\rho_{AB}$. In other words, if a bipartite state of local Hilbert space dimensions $d_A$ and $d_B$ has Schmidt number $K$, we say that it exhibits genuine $K$-dimensional entanglement, i.e.~one requires $K$ internal levels to be entangled in order to generate the state.

\subsection{Schmidt number criterion}
Consider that two distant parties, Alice and Bob, share a high-dimensional bipartite state $\rho_{AB}$ of equal local dimensions $d$. The goal of the parties is to detect entanglement by certifying a lower bound on the Schmidt number of $\rho_{AB}$. While there are many different schemes in the literature for Schmidt number detection, the most practically designed ones are based on performing only a small number of well-selected measurements, as this allows entanglement to be certified with reasonable measurement and data collection resources. 

A particularly well-known criterion for high-dimensional entanglement detection is based on measuring correlations in two complementary bases. The two bases can for instance be selected as the computational basis $\{\ket{j}\}_{j=0}^{d-1}$ and its Fourier transform, $\{\ket{f_j}\}_{j=0}^{d-1}$, where $\ket{f_j} = \frac{1}{\sqrt{d}}\sum_{l=0}^{d-1}\omega^{jl}\ket{l}$ and $\omega = e^{2 \pi i/d}$. The criterion works equally well for any pair of mutually unbiased bases (MUBs), i.e.~the only requirement is that the modulus overlap between any pair elements selected from the two bases is constant, namely $|\langle j | f_k\rangle|^2 = \frac{1}{d}$ for all $j,k$. 
The protocol proceeds as follows. Alice and Bob both measure their respective shares of $\rho_{AB}$ in the computational basis and check if their measurement outcomes are the same. Then, they change their measurements to the bases $\{\ket{f_j}\}_j$ for Alice and to $\{\ket{f_j^*}\}_j$ for Bob, where $\ket{f_j^*}$ is the complex conjugate of $\ket{f_j}$. Again, they check if their measurement outcomes are the same. The sum of the probabilities of their global measurements in which they have the same outcome is labeled $\mathcal{W}_d$. $\mathcal{W}_d$, which we will refer to as entanglement witness, serves as the key parameter from which the entanglement dimensionality is detected. Specifically, as shown first for entanglement\cite{Spengler_2012} and then generalized to the Schmidt number \cite{Morelli_2023}, any state $\rho_{AB}$ with a Schmidt number no larger than $K$ satisfies the inequality 

\begin{equation}\label{eq:SN}
\mathcal{W}_d = \sum_{j=0}^{d-1} \Tr\left(\ketbra{j}{j}  \otimes \ketbra{j}{j}\rho_{AB}\right) +\Tr\left(\ketbra{f_j}{f_j}  \otimes \ketbra{f_j^*}{f_j^*}\rho_{AB}\right)\leq 1 + \frac{K}{d}.
\end{equation}
Thus, a violation of this inequality certifies that $\rho_{AB}$ has a Schmidt number of at least $K+1$. Note, that the  algebraically maximal value of $\mathcal{W}_d$ is 2 and that it can be reached if $\rho_{AB}$ is chosen as the $d$-dimensional maximally entangled state $\ket{\phi^+} = \frac{1}{\sqrt{d}}\sum_{i=0}^{d-1} \ket{i,i}$. 

This high-dimensional entanglement criterion is prevalent due to its favorable features. Firstly, it uses measurements in only two bases, which is the minimal number possible. Secondly, the measurement bases (two MUBs) are natural both from a theoretical point of view as well as accessible in many experimental setups. Thirdly, it closely parallels the high-dimensional generalization of the BB84 protocol for quantum key distribution \cite{Cerf2002, Sheridan2010} which is based on the same local measurements. Fourthly, $\mathcal{W}_d$ allows the experimenter to also infer a bound on the fidelity, $F=\bra{\phi^+}\rho_{AB}\ket{\phi^+}$ between the lab state $\rho_{AB}$ and the ideal maximally entangled state $\phi^+$ through the simple relation $F\geq \mathcal{W}_d-1$ \cite{Morelli_2023}.

\subsection{Measurement imperfections}
In order for the entanglement witness $\mathcal{W}_d$ given in Eq.~\eqref{eq:SN} to be faithfully tested, the measurements implemented in the lab must precisely correspond to those described in the theory, namely the two MUBs.

However, a perfect measurement implementation is an idealization that is increasingly difficult to approximate as quantum systems become more complex. In other words, the fidelity of a measurement on most platforms typically decreases as we increase the Hilbert space dimension. This creates a mismatch between the theoretical pre-requisites of the entanglement witnessing and their lab implementations. 
Since full tomography rapidly becomes too costly in the high-dimensional regime, it is common practice to use the fidelity to benchmark how accurately a lab measurement approximates a target basis. However, there is more than one way in which the fidelity error of a measurement can be quantified. Below, we discuss three separate approaches to quantifying fidelity imperfections in measurements. 

\subsubsection{The average fidelity model}
An intuitive way in which one can quantify how close a set of lab measurement is to the target bases is by considering their average fidelity\cite{Rosset_2012, Cao_2024, Morelli_2022}. That is, the lab measurement can be subject to arbitrary imperfections, but the magnitude of this imperfection causes no more than a bounded fidelity reduction. In our entanglement witness described in Eq.~\eqref{eq:SN}, the target measurements are the computational basis and the Fourier basis respectively. We let $\{E_j\}_j$ and $\{F_j\}_j$ denote the corresponding lab measurements. A bounded average fidelity error means that 
\begin{equation}\label{eq:ave_error}
    \mathcal{F}_{\text{avg}}^{\text{comp}}\equiv \frac{1}{d}\sum_{j=0}^{d-1} \langle j|E_j|j\rangle \geq 1- \varepsilon_E, \qquad \qquad \mathcal{F}_{\text{avg}}^{\text{four}}\equiv\frac{1}{d}\sum_{j=0}^{d-1} \langle f_j|F_j|f_j\rangle  \geq 1- \varepsilon_F.
\end{equation}
Thus, the average infidelity of the lab measurements is bounded by the parameter $\varepsilon \in [0,1]$. When $\varepsilon= 0$, the lab measurements precisely correspond to the target measurements. 
Typically, we are interested in realistic descriptions with a non-zero but small $\varepsilon$. Furthermore, the average fidelity as a clear physical meaning: it is the average probability of obtaining outcome $j$ when the lab measurement is applied to the $j$'th state in the target basis. 

\subsubsection{The worst-case fidelity model}\label{sec:IndepProjector}
A more conservative way to quantify the measurement infidelity is by considering the worst-case fidelity deviation \cite{Tavakoli_2024}. This means that instead of considering the fidelity deviation on average, as in Eq.~\eqref{eq:ave_error}, we bound the fidelity deviation for each possible outcome. Specifically, this means that 
\begin{equation}\label{eq:proj_Error}
    \langle j|E_j|j\rangle \geq 1- \varepsilon_{E}, \qquad \qquad\langle f_j|F_j|f_j\rangle  \geq 1- \varepsilon_{F}, \qquad \forall j .
\end{equation}

\subsubsection{The worst-case fidelity model with individual projections}\label{sec:BinProcess}
When analyzing the effect of measurement imperfections on entanglement witnesses, it is important to consider how the measurements stipulated by theory are realized in practice. 
The witness $\mathcal{W}_d$ assumes local measurements in the computational and Fourier bases, each with $d$ possible outcomes per experimental round. However, due to the difficulty often associated with resolving all $d$ possible outcomes in large dimensions, it is common practice in photonics to replace the $d$-outcome measurement with a sequence of $d$ individual projections. The $j$'th projection in the sequence is associated with the $j$'th outcome of the original measurement. To this end, the computational basis measurement $\{\ketbra{j}{j}\}_j$ is substituted by $d$ separate ``click/no-click'' measurements, $\{E_{\text{\cmark} | j}, E_{\text{\xmark} | j}\}_{j=0}^{d-1}$. In this case, a ``click'' event (\cmark) is intended to represent a successful projection onto the basis vector $\ket{j}$, i.e., $E_{\text{\cmark}|j} = \ketbra{j}{j}$, while a ``no-click'' event (\xmark) its complement $E_{\text{\xmark}|j} = \openone-\ketbra{j}{j}$. Similarly, the Fourier basis is substituted by $\{F_{\text{\cmark} | j}, F_{\text{\xmark} | j}\}_{j=0}^{d-1}$. 

We consider the worst-case fidelity model under such a binarized measurement implementation. The infidelity per individual successful projection becomes
\begin{equation}
    \langle j|E_{\text{\cmark}|j}|j\rangle \geq 1- \varepsilon_{E}, \qquad \qquad\langle f_j|F_{\text{\cmark}|j}|f_j\rangle  \geq 1- \varepsilon_{F} \qquad \forall j,
\end{equation}
where we assume that each successful measurement event $E_{\text{\cmark}|j}$ and $F_{\text{\cmark}|j}$ is a standard rank-one projection.
Importantly, if the individual projections are flawless, then this is equivalent to the multi-outcome description of the measurement. 
However, when $\varepsilon>0$, each projection in the sequence may deviate in a distinct way, implying that the set of $d$ independent projections $\{ E_{\text{\cmark}|j}\}_j$ ($\{ F_{\text{\cmark}|j}\}_j$) does no longer necessarily represent a valid positive operator-valued measure (POVM), (the completeness relation is not satisfies) \cite{Tavakoli_2025}. In this model, this enables different types of deviations in comparison to the one discussed in the previous subsection.

\section{Bounded-fidelity hacking attacks for faking entanglement}\label{App:fake_protocols}
Measurement imperfections may arise not only from stochastic noise and systematic deviations, but also from adversarial influence. 
We present three adversarial attacks on the entanglement witness $\mathcal W_d$, each tailored for one of the three measurement error models described above. 
These attacks are based on (entirely classical) product states and use only $\varepsilon$-sized fidelity reductions of measurement fidelity to falsely certify high-dimensional entanglement.  
While the specific entanglement dimension that can be faked in these attacks depends on the employed error model, we will see that in all cases, a small measurement infidelity is sufficient to cause large false positives in the amount of entanglement. This effect is especially strong in high-dimensional regime, which is many times also the most relevant  regime for high-dimensional quantum information science.

Consider that Alice and Bob share the product state $\rho_{AB} = \psi_A\otimes \psi_B$. Label the lab measurements by $\{E_j\}_j$ and $\{F_j\}_j$, which correspond to the experimenter's attempt at realizing the computational and the Fourier basis, respectively. For simplicity, we assume that both measurements have the same degree of imperfection, $\varepsilon$. To this end, the quantity $\mathcal{W}_d$ used in the entanglement witness in Eq.~\eqref{eq:SN} becomes
\begin{equation}\label{eq:classical_bound}
\begin{aligned}
        \mathcal{W}_d &= \sum_{j=0}^{d-1}  \tr(E_j^A \psi_A) \tr(E_j^*{}^B \psi_B) + \tr( F_j^A \psi_A) \tr(F_j^* {}^B \psi_B) \\ & \leq \sum_{j=0}^{d-1} \sqrt{\tr(E_j^A \psi_A)^2 +\tr(F_j^A \psi_A)^2 } \sqrt{\tr(E_j^*{}^B \psi_B)^2 +\tr(F_j^*{}^B \psi_B)^2 }.
\end{aligned}
\end{equation}
In the first line, we have applied the product state structure and for the second line the Cauchy-Schwarz inequality. Since the target measurements in each product term are identical up to a complex conjugate, we can optimally take $\ket{\psi} \equiv \ket{\psi}_A = \ket{\psi^*}_B$. Moreover, since the measurements of each party only appear independently in the two factors in the second line, we can optimally select them such that the factors are equal. With these choices, the inequality above is saturated. By denoting the witness value achievable with a product state by $\mathcal{W}^{(d)}_{\text{prod}}(\varepsilon)$, we have now arrived at 
\begin{equation}\label{eq:witness_prod}
        \mathcal{W}^{(d)}_{\text{prod}}(\varepsilon) =  \sum_{j=0}^{d-1} \left(\braket{\psi|E_j|\psi}^2+\braket{\psi|F_j|\psi}^2\right).
\end{equation}
On the right-hand side, we are free to choose any state $\ket{\psi}$ and any measurements $\{E_j\}_j$ and $\{F_j\}_j$ that are $\varepsilon$-close to the target bases. Importantly, while $\mathcal{W}^{(d)}_{\text{prod}}(\varepsilon)$ cannot exceed $1+\frac{1}{d}$ for $\varepsilon=0$ due to Eq.~\eqref{eq:SN}, the same does not need be true for any small measurement infidelity satisfying $\varepsilon>0$. If the test is implemented with any non-zero $\varepsilon$, the value $\mathcal{W}^{(d)}_{\text{prod}}(\varepsilon)$ can in principle exceed the bound in Eq.~\eqref{eq:SN}. To an experimenter who is performing a standard entanglement witness test, thereby assuming control of the measurement, this leads per Eq.~\eqref{eq:SN} to 
\begin{equation}\label{eq:rel_SN}
    K =  d(\mathcal{W}^{(d)}_{\text{prod}}(\varepsilon)-1).
\end{equation}
We will refer to the $K$ obtained by this procedure as the faked Schmidt number. Hence, measurement inaccuracy may yield a falsely certified Schmidt number of at least $\lceil K \rceil$.

\subsection{Attack for average fidelity model}\label{sec:average_model}
We start by introducing an attack protocol for measurements that deviate by an $\varepsilon$ quantified by the average fidelity model. 

To this end, we start by building a $d$-dimensional basis $\{\ket{\phi_k}\}_{k=0}^{d-1}$ through the following ansatz,

\begin{equation}\label{eq:meas_ave}
    \begin{aligned}
   &\ket{\phi_{0}} = \alpha \ket{0}+ \beta \sum_{j=1}^{d-1}\ket{j}, \qquad \qquad \ &k=0 \\
       &\ket{\phi_k} = \beta \ket{0} - \gamma \ket{k} + \delta  \sum_{\substack{j = 1\\j \neq k}}^{d-1} \ket{j}  , \quad &k>0.\\ 
    \end{aligned}
\end{equation}
First, for these vectors to form a basis, we require that they are normalized. This fixes two of the coefficients to be

\begin{equation}
    \beta = \sqrt{\frac{1-\alpha^2}{d-1}}, \qquad\qquad \gamma = \sqrt{\frac{\alpha^2+(d-2)(1-(d-1)\delta^2)}{d-1}}.
\end{equation}
Then, we require the vectors to satisfy the orthogonality condition $\braket{\phi_j|\phi_k} = 0$, for all $j\neq k$. This yields two equations in $\alpha$ with several solutions, among which the only relevant one is 

\begin{equation}
    \alpha = 1-(d-1)\delta.
\end{equation}
Next, to optimize the attack, we choose the basis such that it saturates the average fidelity constraint with respect to the computational basis. Solving the associated equation, namely $\frac{1}{d}\sum_{k = 0}^{d-1}\left|\braket{k|\phi_k}\right|^2 = 1- \varepsilon$, we deduce that

\begin{equation}
    \delta = \frac{2}{d} \left( 1- R \right).
\end{equation}
where we have introduced the notation
\begin{equation}\label{eq:R}
    R(\varepsilon,d) \equiv \sqrt{1-\frac{\varepsilon d^2}{4(d-1)}}.
\end{equation}
Hence, the basis $\{\ket{\phi_k}\}_k$ is fully described by 

\begin{equation}
\begin{aligned}
        &\alpha = \frac{2(d-1)R-(d-2)}{d}, \quad \qquad \beta = \frac{2}{d}\sqrt{(1-R)(1+(d-1)R)} \\
        &\gamma = \frac{2R+d-2}{d},\qquad \qquad \qquad \quad   \delta = \frac{2}{d}(1-R).
\end{aligned}
\end{equation}
Note, that this solution is valid only when $R$ is real. This corresponds to selecting $\varepsilon$ to be below the critical value $\varepsilon_{\text{crit}} = 4(d-1)/d^2$. Although this threshold monotonically decreases as a function of the dimension, it remains relevant for high dimensions for the small-deviation regime that we are interested in. For example, at $d =100$, the critical value equals $\varepsilon_{\text{crit}} = 3.96 \%$.

We now select the lab measurements associated with the computational basis and its Fourier basis as $E_k = \ketbra{\phi_k}{\phi_k}$ and $F_k = G \ketbra{\phi_k}{\phi_k} G^\dagger$, respectively, where $G$ is the Fourier matrix 
\begin{equation}
    G = \frac{1}{\sqrt{d}} \sum_{j,k=0}^{d-1} \omega^{jk}\ketbra{j}{k},
\end{equation}
with $\omega = e^{2 \pi i/d}$. Clearly, the perturbed Fourier basis saturates the average fidelity constraint, since $\sum_k\langle f_k |F_k|f_k\rangle = \sum_k\langle k| G^\dagger G E_k G^\dagger G |k\rangle = \sum_k\langle k|E_k|k\rangle = 1-\varepsilon$.

Let us now focus on the selection of the state $\ket{\psi}$. To that end, following the above discussion, we write the witness value for product states as 
\begin{equation}\label{eq:bound_mod1}
             \mathcal{W}^{(d)}_{\text{prod}}(\varepsilon)=  \sum_{k=0}^{d-1} \left(\lvert\braket{\psi|\phi_k}\rvert^4 + \lvert\braket{\psi|G|\phi_k}\rvert^4 \right).
\end{equation}
We now seek to optimize $\mathcal{W}^{(d)}_{\text{prod}}(\varepsilon)$ over $\ket{\psi}$. To this end, consider the ansatz
\begin{equation}\label{stateansatz}
    \ket{\psi} =  \nu\ket{0}+\mu \sum_{j=1}^{d-1}\ket{j},
\end{equation}
where $\nu = \sqrt{1-(d-1)\mu^2}$ is given by the normalization. In order to maximize the witness value, we want to solve the equation $\frac{d \mathcal{W}^{(d)}_{\text{prod}}(\varepsilon)}{d \mu} = 0$. 
In general, we are unable to analytically solve this equation. 
However, by numerically optimizing the function $\mathcal{W}^{(d)}_{\text{sep}}(\varepsilon)$ over $\mu$ with a fixed $\varepsilon$ and $d \in [2,100]$, we observe that above a critical infidelity, we can find the relevant analytical solution for any choice of $d$. This solution corresponds to
\begin{equation}\label{eq:mu}
    \mu = \frac{1}{\sqrt{2}\sqrt{d+\sqrt{d}}}, \qquad\qquad   \nu = \frac{1}{\sqrt{2}}\sqrt{1+\frac{1}{\sqrt{d}}}.
\end{equation}
We display the critical infidelity $\tilde \varepsilon_{\text{crit}}$ as a function of the dimension in Fig.~\ref{fig:critImprAve}. There, we  see that the critical value is always small, showing that one can reliably use the above analytical solution. The highest critical value $\tilde \varepsilon_{\text{crit}} \approx 3 \times 10^{-3}$ is achieved with $d = 11$. In view of this, we proceed using the state parameters in Eq.~\eqref{eq:mu}.

\begin{figure}[ht!]
    \centering
    \includegraphics[width=0.5\linewidth]{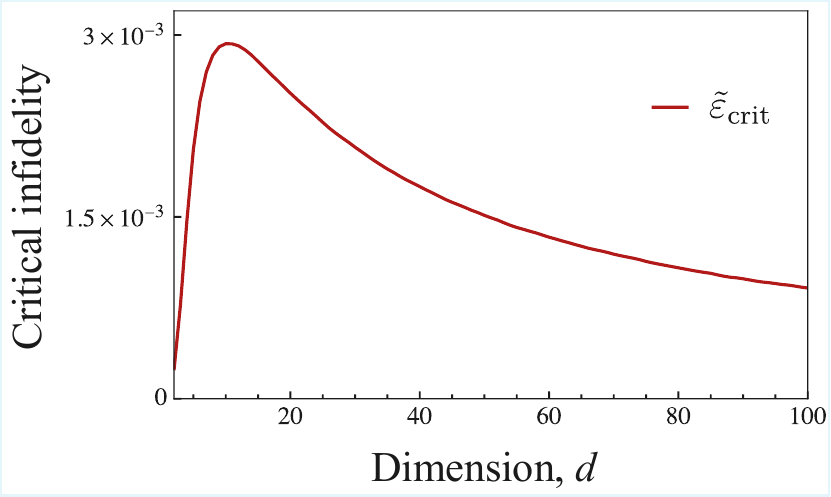}
    \caption{Critical infidelity as a function of dimension, where $\mu$ can be chosen according to Eq.~\eqref{eq:mu}. }
    \label{fig:critImprAve}
\end{figure}

With all parameters fixed, we now evaluate the witness parameter. For this purpose, we compute the overlaps between the product state and the lab measurements appearing in Eq.~\eqref{eq:bound_mod1}. In the first basis the overlaps are given by
\begin{equation}\label{eq:overlaps_ave_comp}
\braket{\psi|\phi_k} = 
    \begin{cases}
        & \alpha \nu+(d-1)\mu \beta , \qquad \qquad k = 0 \\
        &\beta \nu + \mu \left((d-2)\delta - \gamma \right), \quad \ \ k >0.\\
    \end{cases}
\end{equation} 
In the second basis the overlaps are
\begin{equation}\label{eq:overlaps_ave_four}
\sqrt{d}\braket{\psi|G|\phi_k} = 
    \begin{cases}
            & \nu (\alpha+(d-1)\beta)+ \mu(d-1)(\alpha-\beta), \qquad \qquad \qquad \quad  \ \ k = 0\\
        & \nu(\beta-\gamma+(d-2)\delta)+\mu((d-1)(\beta-\delta) + \gamma+\delta) , \qquad k > 0.
    \end{cases}
\end{equation} 
Inserting these equations into Eq.~\eqref{eq:bound_mod1}, the entanglement witness parameter becomes
\begin{equation}
\begin{aligned}
         \mathcal{W}^{(d)}_{\text{prod}}(\varepsilon) &= \left|\braket{\psi|\phi_{0}}\right|^4+ \left|\braket{\psi|G |\phi_{0}}\right|^4+ (d-1) \left( \lvert \braket{\psi|\phi_1}\rvert^4+ \left|\braket{\psi|G|\phi_1}\right|^4\right).
\end{aligned}
\end{equation}
By simplifying the expression, we obtain \begin{equation}
    \mathcal{W}^{(d)}_{\text{prod}}(\varepsilon) = \frac{(\sqrt{d}+1)^3\lvert  \alpha+(\sqrt{d}-1)  \beta \rvert^4+(\sqrt{d}-1)\lvert \alpha -(\sqrt{d}+1)  \beta \rvert^4}{2d (\sqrt{d}+1)}.
\end{equation}

We note, that this formula applies only when $\varepsilon$ exceeds the threshold in Fig.~\ref{fig:critImprAve} (which is no larger than $3\times 10^{-3}$). However, for any dimension and any $\varepsilon$ below this threshold, computing the value of $\mathcal{W}^{(d)}_{\text{prod}}(\varepsilon)$ is straightforward since we have reduced the problem to an optimization over a single real-valued parameter, namely $\mu$, as defined in Eq.~\eqref{stateansatz}. Hence, although we do not have a closed form of $\mathcal{W}^{(d)}_{\text{prod}}(\varepsilon)$ below the threshold in $\varepsilon$, it is easily computed in any given case.

It is relevant to ask whether the attack outlined here is the best one possible. In section~\ref{App:Performance} we provide evidence that the answer is positive. 

\subsection{Attack for worst-case fidelity model}\label{sec:worst_case}
In this section, we develop an attack protocol for measurements that deviate by an $\varepsilon$ quantified by the worst-case fidelity model.

Let the product state take the same form as in the attack for the average fidelity model. The local state therefore reads
\begin{equation}
    \ket{\psi} =  \nu\ket{0}+\mu \sum_{j=1}^{d-1}\ket{j},
\end{equation}
where $\nu = \sqrt{1-(d-1)\mu^2}$. 

We denote $E_k$ and $F_k$ as the lab measurements associated with the computational basis and its Fourier transform, respectively, with $k = 0,...,d-1$. In analogy to the average fidelity model, we let the $k$'th element in the former and latter set be related through the Fourier matrix $G$, such that $F_k= G E_k G^\dagger$ for all $k$. We now construct the POVM $\{E_k\}_k$. To this end, take the first POVM element to be a rank-two operator, and the remaining elements to be sub-normalized rank-one projectors 
\begin{equation}\label{eq:meas_worst}
\begin{aligned}
 &E_0 =\ketbra*{\phi_0^\uparrow}{\phi_0^\uparrow} + \lambda_1 \ketbra*{\phi_0^\downarrow}{\phi_0^\downarrow}, \qquad \qquad k=0\\
    &E_k = \lambda_2 \ketbra{\phi_k}{\phi_k}, \qquad \qquad \qquad\qquad \ \ k>0.
\end{aligned}
\end{equation}
Our ansatz for these vectors reads

\begin{equation}
\begin{aligned}
&\ket{\phi_0^\uparrow} = \alpha_1 \ket{0} + \beta_1 \sum_{j=1}^{d-1}\ket{j} ,  &k=0 \\
&\ket{\phi_0^\downarrow} =  - \alpha_2 \ket{0} + \beta_2 \sum_{j=1}^{d-1}\ket{j},  &k=0\\
&\ket{\phi_k} = \gamma \ket{0} - \delta \ket{k} +\xi \sum_{\substack{j=1 \\ j \neq k}}^{d-1}\ket{j},  &\qquad k>0.
\end{aligned}
\end{equation}
We now determine the coefficients as functions of the dimension $d$ and the infidelity parameter $\varepsilon$. From normalization, we can directly deduce $(\alpha_1,\alpha_2,\gamma)$ to be
\begin{equation}
    \begin{aligned}
    &\alpha_1 = \sqrt{1-(d-1)\beta_1^2}, \\
    &\alpha_2 = \sqrt{1-(d-1)\beta_2^2}, \\ 
    &\gamma=\sqrt{1-\delta^2-(d-2)\xi^2}.
\end{aligned}
\end{equation}
Moreover, for the elements to form a valid POVM, we require that $\ket{\phi_0^\uparrow}$ and $\ket{\phi_0^\downarrow}$ are orthogonal. From the condition $\braket{\phi_0^\uparrow|\phi_0^\downarrow} = 0$, the parameter $\beta_2$ is deduced as 
\begin{equation}
    \beta_2 = \sqrt{\frac{1}{d-1}-\beta_1^2}.
\end{equation}
Similarly, to form a valid POVM we require that each $\ket{\phi_k}$ is orthogonal to $\ket{\phi_0^\uparrow}$ for $k>0$. This gives
\begin{equation}
    \beta_1 = \sqrt{\frac{1-\delta^2-(d-2)\xi^2}{(d-1)-(d-2)(\delta+\xi)^2}}.
\end{equation}

Next we impose that the measurement elements satisfy the completeness relation $\sum_k E_k = \openone$. By tracing both sides of the completeness relation we obtain an equation which we can solve for $\lambda_1$, namely $1+\lambda_1 + (d-1)\lambda_2 = d$. Then, projecting the completeness relation onto the computational basis vector $\ket{0}$, we obtain the following equation: $1 = \langle 0 |(\sum_k E_k)|0\rangle=\alpha_1^2+\lambda_1 \alpha_2^2 + \lambda_2(d-1) \gamma^2 $, which we can solve for $\lambda_2$. From these two equations, we deduce that
\begin{equation}
    \lambda_1 = (d-1)\left(1-\frac{1}{(\delta+\xi)^2}\right),\qquad \qquad \lambda_2 = \frac{1}{(\delta+\xi)^2}.
\end{equation}

The remaining free parameters are $\delta$ and $\xi$. These are obtained by saturating the worst-case fidelity constraints $\braket{k|E_k|k}= 1- \varepsilon$. Note that $\braket{0|E_0|0} =  \alpha_1^2 + \lambda_1 \alpha_2^2$ for $k=0$, whereas $\braket{k|E_k|k} = \lambda_2 \delta^2$ for all $k >0$. Hence, it is sufficient to solve $\braket{0|E_0|0}= 1- \varepsilon$ and $\braket{1|E_1|1}= 1- \varepsilon$. Doing so, we find that
\begin{equation}
    \begin{aligned}
    &\delta =  \left(1+\sqrt{1-\varepsilon}\right)\frac{ \sqrt{(d-1)(1-\varepsilon)}}{R} \\
    &\xi = \frac{\varepsilon \sqrt{d-1}}{R},
    \end{aligned}
\end{equation}
where we have defined
\begin{align}
    R(\varepsilon,d) \equiv 2(d-1)+2(d-1-(d-2)\varepsilon)\sqrt{1-\varepsilon}+(5-3d+(d-2)d\varepsilon).
\end{align}
This completes the description of the POVM $\{E_k\}_k$. Note that all $F_k$ satisfy the fidelity constraint, since $\langle f_k|F_k|f_k\rangle = \langle k|E_k|k\rangle = 1-\varepsilon$ for all elements $k$. 

\medskip

The ansatz for the POVM $\{E_k\}_k$ is valid up to a critical infidelity. This is seen by studying the completeness relation $\sum_k E_k = \openone$. For this relation to be satisfied we require that $0 = \langle 0 | \sum_k E_k |j\rangle = \alpha_1\beta_1 - \lambda_1 \alpha_2 \beta_2 + \lambda_2 \gamma((d-2)\xi-\delta)$ for $j>0$. This relations holds when $(d-2)\xi-\delta \leq 0$. Solving the equality for $\varepsilon$, the critical infidelity is found to be
\begin{equation}
    \varepsilon_{\text{crit}} = \frac{2d-3}{(d-1)^2}.
\end{equation}
Below this threshold the ansatz is valid and the completeness condition is satisfied. Although the critical infidelity monotonically decreases as a function of the dimension, it remains relevant for the small-deviation regime that we are interested in. For example, at $d =100$, the critical value equals $\varepsilon_{\text{crit}} = 2.01 \%$. Furthermore, as we will soon see, in the range $d \in [2,100]$ the witness parameter always reaches the algebraic bound $ \mathcal{W}^{(d)}_{\text{prod}}(\varepsilon) = 2$ with an infidelity smaller than $\varepsilon_{\text{crit}}$. 

\medskip
We now express all parameters in the attack protocol in terms of the dimension ($d$) and the infidelity ($\varepsilon$):

\begin{equation}
    \begin{aligned}
            &\alpha_1 = \sqrt{1-\frac{Y(d-1)}{X}}, \ \  \alpha_2 = \sqrt{\frac{Y(d-1)}{X}}, \ \ \beta_1 = \sqrt{\frac{Y}{X}}, \ \ \beta_2 = \sqrt{\frac{1}{d-1}-\frac{Y}{X}}, \\[1em]
        &\gamma = \sqrt{Y}, \qquad \delta = (1+\sqrt{1-\varepsilon})\sqrt{\frac{(d-1)(\varepsilon-1)}{R}}, \qquad \xi = \varepsilon \sqrt{\frac{d-1}{R}}, \\[1em]
        &\lambda_1 = (d-1)-\frac{R}{ (1 + \sqrt{1 - \varepsilon})^2} , \qquad \lambda_2 = \frac{R}{(d-1) (1 + \sqrt{1 - \varepsilon})^2},
    \end{aligned}
\end{equation}
where we have for simplicity defined
\begin{equation}
    \begin{aligned}
        X = (d-1)-(d-2)(\delta+\xi)^2, \qquad \qquad Y= 1-\delta^2-(d-2)\xi^2.
    \end{aligned}
\end{equation}

\medskip

\begin{figure}[t!]
    \centering
    \includegraphics[width=0.5\linewidth]{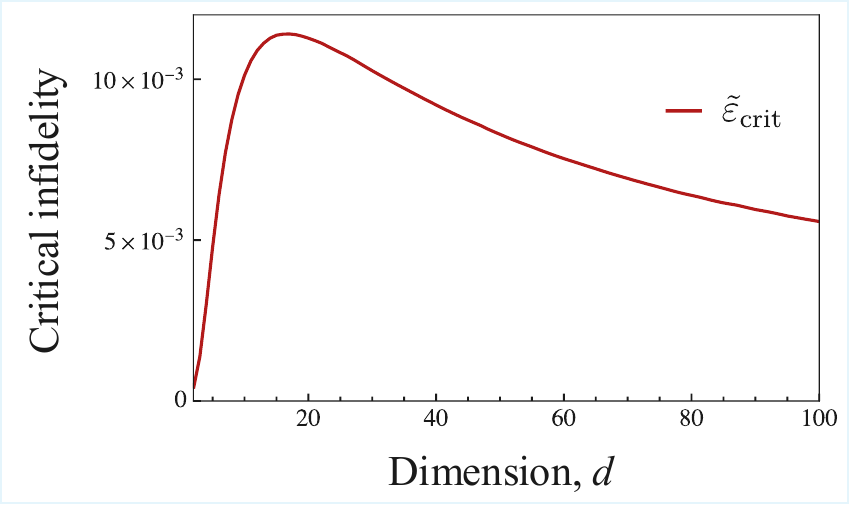}
    \caption{The critical infidelity as a function of dimension. For $\varepsilon$ above this value, the optimal parameter choice is $\mu =\frac{1}{\sqrt{2}\sqrt{d+\sqrt{d}}}$.}
    \label{fig:critImprMulti}
\end{figure}

We now evaluate the witness parameter 
\begin{equation}
    \mathcal{W}_{\text{prod}}^d(\varepsilon)=  \sum_{k=0}^{d-1} \left(\left|\braket{\psi|E_k|\psi}\right|^2 +\lvert\braket{\psi|G E_k G^\dagger|\psi}\rvert^2\right).
\end{equation}
Expanding the expression we find that
\begin{equation}\label{eq:bound_mod2}
\begin{aligned}
 \mathcal{W}^{(d)}_{\text{prod}}(\varepsilon) & =\lvert \braket{\psi|\phi_{0}^\uparrow}\rvert^4+\lvert\braket{\psi|G | \phi_{0}^\uparrow}\rvert^4 + \lambda_1^2 \big( \lvert\braket{\psi|\phi_{0}^\downarrow}\rvert^4+\lvert \braket{\psi|G| \phi_{0}^\downarrow}\rvert^4 \big)  \\
 &+ 2\lambda_1 \big(\lvert\braket{\psi|\phi_{0}^\uparrow}\rvert^2  \lvert\braket{\psi|\phi_{0}^\downarrow}\rvert^2
+\lvert\braket{\psi|G | \phi_{0}^\uparrow}\rvert^2  \lvert\braket{\psi|G| \phi_{0}^\downarrow}\rvert^2 \big)\\&+ \lambda_2^2 (d-1)\big(\lvert\braket{\psi|\phi_{1}} \rvert^4+  \lvert\braket{\psi|G |\phi_{1}}\rvert^4\big) .
\end{aligned}
\end{equation}
Computing the overlaps between $\ket{\psi}$ and the rays of the first POVM we find

\begin{equation}\label{eq:wc_overlap_comp}
\braket{\psi|\phi_k}  = 
\begin{cases}
&\nu \alpha_1+ (d-1)\mu\beta_1 \qquad \qquad   k = 0 \text{ and } \uparrow \\
& - \nu \alpha_2 + (d-1)\mu\beta_2 \qquad \quad \ k = 0 \text{ and } \downarrow \\
& \nu \gamma - \delta \mu + (d-2)\xi \mu \ \ \ \quad \quad k >0.
\end{cases}
\end{equation}
Similarly, the overlaps between $\ket{\psi}$ and Fourier transformed rays read
\begin{equation}\label{eq:wc_overlap_four}
\sqrt{d} \langle \psi| G |\phi_k\rangle  = 
\begin{cases}
& \nu(\alpha_1+\beta_1(d-1)) + \mu(\alpha_1-\beta_1)(d-1),  \ \qquad \quad \ k = 0 \text{ and } \uparrow \\
&  \nu(-\alpha_2+\beta_2(d-1)) - \mu(\alpha_2+\beta_2)(d-1),  \qquad  \ \ \ k = 0 \text{ and } \downarrow \\
& \nu (\gamma-\delta+\xi(d-2)) + \mu((d-1)(\gamma-\xi)+\delta +\xi), \quad  k >0.
\end{cases}
\end{equation}

The only free parameter left in the attack protocol is $\mu$. In general, we cannot solve for $\mu$ analytically. However, the choice $\mu =1/(\sqrt{2}\sqrt{d+\sqrt{d}})$ provides a solution that is optimal above some critical infidelity $\tilde \varepsilon_{\text{crit}}$ that depends on the dimension; it is illustrated in Fig.~\ref{fig:critImprMulti}. Below this threshold, the value of $\mathcal{W}_{\text{prod}}^d(\varepsilon)$ is easily computed for any given $d$ and $\varepsilon$ by performing an optimization over the single real parameter $\mu$.

In section \ref{App:Performance} we provide positive evidence that the attack outlined here is the best possible. 

\subsection{Attack for worst-case fidelity model with individual projections}\label{sec:Bin_worst_case}
In this section, we develop an attack protocol for individual projections that deviate by an $\varepsilon$ quantified through the worst-case fidelity model. Denote by $E_k \equiv E_{\text{\cmark}|k}$ and $F_k \equiv  F_{\text{\cmark}|k}$ the $d$ individual rank-one projections associated with successful detection event (``click''). The key difference of this and the previous scenario is that for $\varepsilon >0$, the sets $\{E_k\}_k$ and $\{F_k\}_k$ do not have to form valid measurements (i.e.~they do not satisfy the completeness condition) because the small deviations can be applied independently to each projector in the target basis. 

To construct the attack, we again let the local state take the form
\begin{equation}
\ket{\psi} = \nu \ket{0} + \mu \sum_{j=1}^{d-1}\ket{j},
\end{equation}
where the parameter $\nu = \sqrt{1-(d-1)\mu^2}$ is fixed from normalization. 

We proceed with constructing the set of individual projectors. To this end, we take $E_k = \ketbra{\phi_k}{\phi_k}$, where the associated rank-one projectors reads
\begin{equation}
\begin{aligned}
&\ket{\phi_0} = \alpha \ket{0} +\beta \sum_{j=1}^{d-1} \ket{j}, \qquad \qquad k = 0 \\
&\ket{\phi_k} = \gamma \ket{0} + \alpha \ket{k} + \delta \sum_{\substack{j=1\\j\neq k}}^{d-1} \ket{j}, \quad k>0.
\end{aligned}
\end{equation}
Similarly, we take $F_k = G\ketbra{\varphi_k}{\varphi_k}G^\dagger$, where $G$ is the Fourier matrix, and the rays take the form
\begin{equation}
\begin{aligned}
&\ket{\varphi_0} = \alpha \ket{0} +\beta \sum_{j=1}^{d-1}\ket{j}, \qquad \qquad k = 0 \\
&\ket{\varphi_k} = \xi \ket{0} + \alpha \ket{k} + \eta \sum_{\substack{j=1\\j\neq k}}^{d-1} \ket{j}, \quad k>0.
\end{aligned}
\end{equation}
The coefficients obey $\alpha,\beta,\gamma,\delta,\xi,\eta \geq 0$. From normalization, we directly deduce that
\begin{equation}
    \beta = \sqrt{\frac{1-\alpha^2}{d-1}}, \qquad \delta = \sqrt{\frac{1-\alpha^2-\gamma^2}{d-2}}, \qquad \eta = \sqrt{\frac{1-\alpha^2-\xi^2}{d-2}}.
\end{equation}
Next, we choose the projectors such that they saturate the worst-case fidelity constraint with respect to the target measurements, i.e., $\braket{k|E_k|k} = 1-\varepsilon$ and $\braket{f_k| F_k|f_k} = 1-\varepsilon$, respectively. From this, the parameter $\alpha$ can be determined to be
\begin{equation}
    \alpha = \sqrt{1-\varepsilon}.
\end{equation}

We proceed with computing the witness parameter $\mathcal{W}^{(d)}_{\text{prod}}(\varepsilon)= \sum_{k=0}^{d-1} \left(\left|\braket{\psi|\phi_k}\right|^4 +\left|\braket{\psi|G | \varphi_k}\right|^4\right)$. For the first set of projectors the overlaps are
\begin{equation}
\braket{\psi|\phi_k}=
\begin{cases} 
&\nu\sqrt{1-\varepsilon} + \sqrt{1-\nu^2}\sqrt{\varepsilon}, \qquad \qquad \qquad \ \ \qquad k = 0\\
&\gamma \nu + \mu\big(\sqrt{1-\varepsilon^2} +  \sqrt{(d-2)(\varepsilon-\gamma^2)}\big), \qquad k >0.
\end{cases}
\end{equation}
For the second set of projectors the overlaps are
\begin{equation}
\sqrt{d}\braket{\psi|G|\varphi_k} = 
\begin{cases} 
& \sqrt{1-\varepsilon}(\nu + (d-1)\mu) + \sqrt{\varepsilon} \sqrt{(d-1)}(\nu - \mu), \qquad  \quad k = 0\\ \\
&\left(\sqrt{1-\varepsilon}+\sqrt{(d-2)(\varepsilon-\xi^2)}\right)(\nu-\mu) \\
&\quad + \xi(\nu+(d-1)\mu) , \qquad \qquad \qquad \qquad \qquad \qquad \quad k>0.
\end{cases}
\end{equation}
The witness parameter then reads 
\begin{equation}
    \mathcal{W}^{(d)}_{\text{prod}}(\varepsilon)=    \lvert\braket{\psi|\phi_0}\rvert^4 +(d-1)\lvert\braket{\psi|\phi_1}\rvert^4 +\lvert\braket{\psi|G | \varphi_0}\rvert^4+(d-1)+\lvert\braket{\psi|G | \varphi_1}\rvert^4.
\end{equation}
This expression presently depends on the parameters $(\gamma,\xi,\mu,d,\varepsilon)$. We now seek to optimize the witness with respect to the free parameters $(\gamma,\xi)$. This amounts to solving the two independent equations $\frac{\partial  \mathcal{W}_{\text{prod}}^{(d)}(\varepsilon)}{\partial \gamma}  =  0$ and $\frac{\partial  \mathcal{W}_{\text{prod}}^{(d)}(\varepsilon)}{\partial \xi}  = 0$. The solution to the former and latter equation is given by
\begin{equation}
\begin{aligned}
&\gamma = \mu \sqrt{\frac{\varepsilon }{1-\nu^2}},\\
&\xi =  (\mu +(d-1)\nu) \sqrt{\frac{\varepsilon}{(d-1)+(d-2)\nu^2+2\mu\nu }}.
\end{aligned}
\end{equation}

Using this, the overlaps become
\begin{equation}
\begin{aligned}
        &\braket{\psi|\phi_0} = \sqrt{1 - \varepsilon} \nu + \sqrt{\varepsilon} \sqrt{1 - \nu^2},\\
        &\braket{\psi|\phi_1} = \frac{\sqrt{1 - \varepsilon} \sqrt{1 - \nu^2} + \sqrt{\varepsilon} \sqrt{d-2 +\nu^2}}{\sqrt{d-1}},\\
         &\braket{\psi|G | \varphi_0} =  \frac{\sqrt{\varepsilon} (\sqrt{d-1} \nu - \sqrt{1 - \nu^2}) + \sqrt{1 - \varepsilon} (\mu + 
    \sqrt{d-1} \sqrt{1 - \nu^2})}{\sqrt{d}},\\
         &\braket{\psi|G | \varphi_1} = \bigg( 
 \sqrt{\frac{\varepsilon}{
  d (d - 1) - 
   R}} \left(\sqrt{d-1} + (\nu + (d-1) \mu)^2 + (d-2) (\nu- \mu) \sqrt{R}\right) \\
   &\qquad \qquad \qquad \qquad +\sqrt{1 - \varepsilon} (\nu - \mu)\bigg)\frac{1}{\sqrt{d}}.
\end{aligned}
\end{equation}
Here, we have introduced $R  \equiv 1+(d-2)\nu^2-2(d-1)\nu \mu$ for simpler notation. While we have been unable to analytically determine the last free parameter, $\mu$, we can easily and accurately compute $\mathcal{W}_{\text{prod}}^{(d)}(\varepsilon)$ for any given $d$ and $\varepsilon$ as an optimization over the single real parameter $\mu$. 

In section~\ref{App:Performance} we provide evidence that the attack outlined here is the best possible. 

\section{Performance of the attacks}\label{App:Performance}
In this section, we discuss the performance of the three attacks developed in section~\ref{App:fake_protocols}. We first present numerical evidence that all three attacks are optimal within their respective infidelity frameworks. Then, we discuss their performance in exploiting the small measurement deviations to falsely indicate the presence of entanglement.

\subsection{Evidence for optimality}
We now assess the optimality of our three attack protocols. For this, we compare the values of $\mathcal{W}^d_{\text{prod}}(\varepsilon)$ achieved in our protocols with those obtained from numerical optimization over all possible attacks. These optimizations are implemented using an alternating convex search heuristic based on semi-definite programming routines \cite{sdpreview}.

Specifically, the search optimizes the witness value over all product states and all lab measurements satisfying the fidelity constraints. The routine is decomposed into three sub-routines: 
\begin{enumerate}
    \item optimization over the measurements on system A,\\

    \item optimization over the measurements on system B,\\

    \item optimization over the shared product state $\rho_{AB} = \psi_A \otimes \psi_B$.
\end{enumerate}
These three routines are iterated until the desired convergence is achieved. The first two optimizations can be cast as semidefinite programs. Since the fidelity constraints between the lab and target measurements are linear constraints, these are easily built into the semidefinite program. However, for the third model, we cannot directly impose in the semidefinite program that each of the individual projectors should be of rank one. Instead, we relax the rank-one projective measurement to a unit-trace measurement. While this means that the optimization is performed over a superset of the allowed strategies, we always observe that the final strategy returned by the heuristic optimization procedure corresponds to a valid, rank-one measurement strategy. Moreover, the third optimization step is itself implemented as an alternating convex search over $\psi_A$ and $\psi_B$. Thus, we use an iterative heuristic search as a sub-routine in the main iterative heuristic search. By implementing step three in this way, each optimization over $\psi_A$ and $\psi_B$, respectively, can be implemented as a largest-eigenvalue problem. This way, we systematically search for the largest achievable value of $\mathcal{W}^d_{\text{prod}}(\varepsilon)$.

The described search procedure is only efficient in low dimensions. Hence, for all the three models, we have restricted the search to dimensions $d\in\{2,3,\ldots,20\}$. In the relevant small-$\varepsilon$ regime, our brute-force numerical search for each protocol systematically returns the same value of $\mathcal{W}^d_{\text{prod}}(\varepsilon)$ that we obtained in the protocols detailed in sections~\ref{App:fake_protocols}. This suggests that all three attack protocols are optimal with respect to the given constraints, i.e., that $\mathcal{W}^{(d)}_{\text{prod}}(\varepsilon)$ corresponds to the largest witness value achievable when using a $d$-dimensional product state and measurements with infidelity $\varepsilon$.

\subsection{Faking entanglement}
Each of the three distinct attack protocols is able to fake high-dimensional entanglement from a state that is fully classical (product state) by introducing small adversarial deviations in the measurements. We can benchmark the amount of false entanglement by the Schmidt number $K$ associated with the measured witness value, $\mathcal{W}^{(d)}_{\text{prod}}(\varepsilon)$, as given in Eq.~\eqref{eq:rel_SN}. In Fig.~\ref{fig:CompModelsK} we show, for each of the three attack protocols (solid lines), the inferred Schmidt number as a function of the dimension. Each plot represents a different choice of infidelity, selected as $\varepsilon \in \{ 0.7 \%, 1\%, 2\%, 5\%\}$. The Schmidt number of the maximally entangled state $K=d$ (gray dashed line) is included as a reference for comparison. Furthermore, in Fig.~\ref{fig:CompModelDim} we show the inferred Schmidt number, for each of the three attack protocols (solid lines), as a function of infidelity, for some fixed dimensions $d \in \{10,25,50,75,100\}$. 

Consider first Figs.~\ref{fig:CompModelsK}\,a-c). For low dimensions, the attack based on the average fidelity model (black) is able to fake larger Schmidt numbers than the attacks based on the worst-case fidelity model (blue and red). This is expected, since the latter fidelity model, in contrast to the former, imposes constraints on the level of each individual measurement operator. Next, we observe that the attack based on individual rank-one projectors (blue) outperforms the attack based on general measurements (red) in low dimensions, namely for $d \leq 39$ at $\varepsilon = 0.7 \%$, $d \leq 25$ at $\varepsilon = 1\%$, and $d \leq 15$ at $\varepsilon = 2\%$. Above these dimensions, however, the latter attack (red) performs better. This suggests that attacks which exploit higher-rank measurement operators can enable larger amounts of false entanglement than attacks based on individual rank-one projections. Their advantage becomes more pronounced as the infidelity increases.

The two attack protocols based on general measurements (black and red) are only shown up to the critical dimension, at which their faked Schmidt number saturates $K = d$, i.e.~the value achieved by the maximally entangled state in the relevant dimension. Therefore, above this dimension, the models cease to be relevant since they already have achieved the algebraically largest possible amount of faked entanglement, i.e.~they cannot achieve even larger values of $\mathcal{W}^{(d)}_{\text{prod}}$. This stands in contrast to the individual projections model (blue), where the value of $\mathcal{W}^{(d)}_{\text{prod}}$ continues to grow after reaching $\mathcal{W}^{(d)}_{\text{prod}}=2$, thereby indicating Schmidt numbers larger than $K=d$. This behavior is characteristic for attacks based on a set of individual projectors, as the possibility to influence each individual projector independently no longer ensures that their collection forms a valid quantum measurement\cite{Tavakoli_2025}. Consequently, for large enough infidelity or dimension, these models are able to yield stronger correlations than those physically meaningful under the original assumptions of the entanglement test. This can clearly be seen for $\varepsilon = 5\%$ in Fig.~\ref{fig:CompModelsK}\,d) and would also be visible in Figs.~\ref{fig:CompModelsK}\,a-c) if the dimension axis was extended further.

Lastly, we consider the dependence of the faked Schmidt number of the infidelity, as shown in Fig.~\ref{fig:CompModelDim}. We observe, for each of the three attacks, that the infidelity needed to fake a given Schmidt number decreases when increasing the dimension. This implies, that the entanglement witness test become more vulnerable to adversarial attacks in large dimensions. Moreover, for the average and worst-case fidelity models, Figs.~\ref{fig:CompModelDim}\,a-b), we see that the faking capability only remain modest for very small $\varepsilon$, but then rapidly increase with $\varepsilon$ until the associated Schmidt number reaches its largest possible value $K=d$. The latter regime is the one that is more practically relevant. For example, at just $1\%$ infidelity in dimension $d=100$, we can fake $K=100$ and $K=74$ in the two models respectively. The behavior in the worst-case fidelity model with individual projections, Fig.~\ref{fig:CompModelDim}\,c), is somewhat different. While larger dimension remain more sensitivity to attacks, we observe an initially more steady growth of the Schmidt number, which then increases more rapidly until it even exceeds the algebraic limit of the original entanglement test (dashed lines).

\newpage 
\begin{figure}[ht!]
    \makebox[\linewidth][c]{%
        \includegraphics[width=1\linewidth]{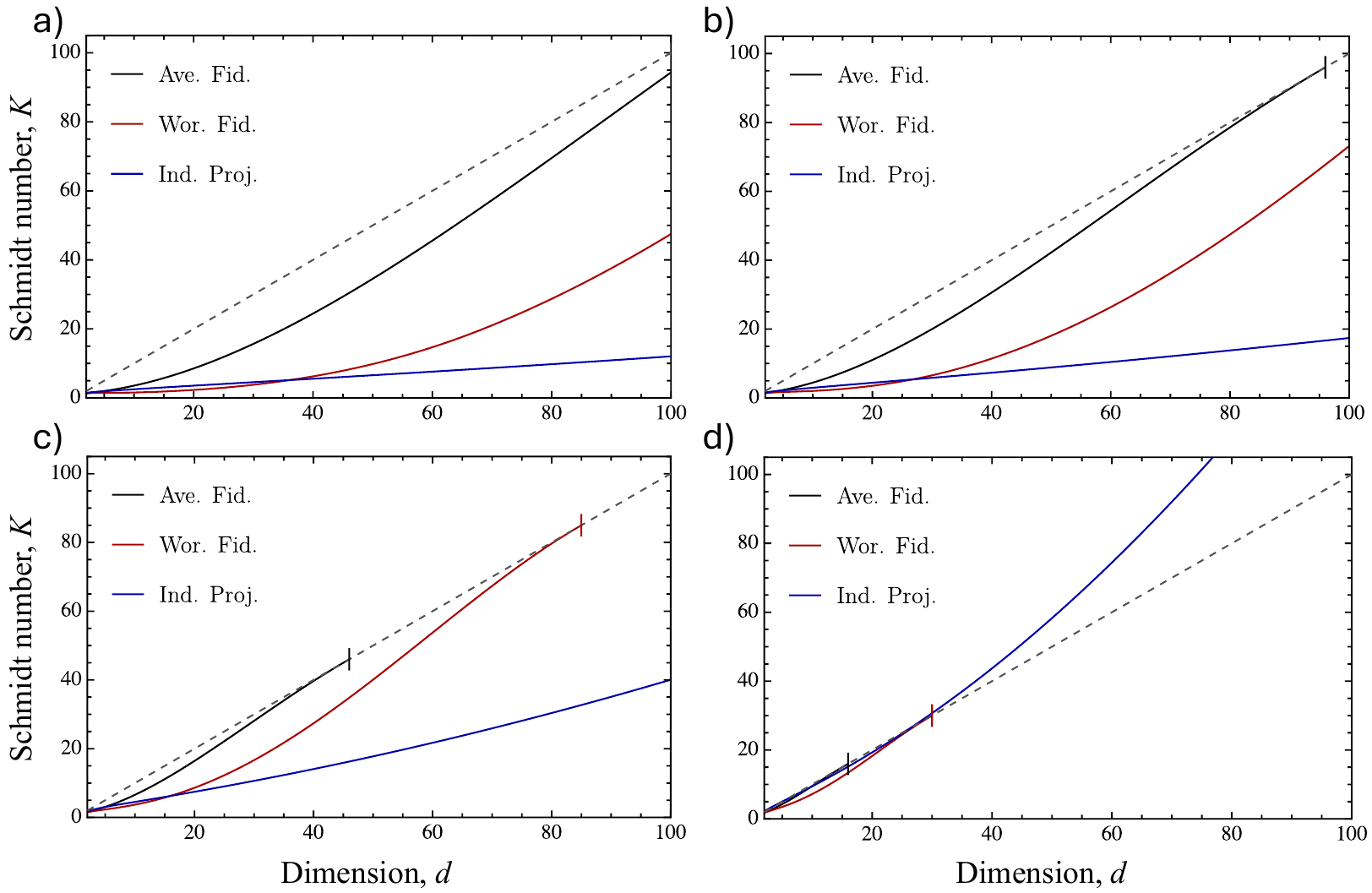}
    }
    \captionsetup{width=1\linewidth}
    \caption{False Schmidt number as a function of dimension for fixed infidelity $\varepsilon$ for an attack based on: (i) average fidelity deviations (black), (ii) worst-case fidelity deviations (red), and (iii) worst-case fidelity deviations with individual projections (blue). The gray dashed line corresponds to the largest meaningful Schmidt number ($K=d$) under the idealized measurement assumptions. The cut-off at which models (i) and (ii) reach $K = d$ is marked. The infidelity is fixed to a) $\varepsilon = 0.007$, b) $\varepsilon = 0.01$, c) $\varepsilon = 0.02$, and d) $\varepsilon = 0.05$.}
    \label{fig:CompModelsK}
\end{figure}

\begin{figure}[ht!]
    \makebox[\linewidth][c]{%
        \includegraphics[width=1\linewidth]{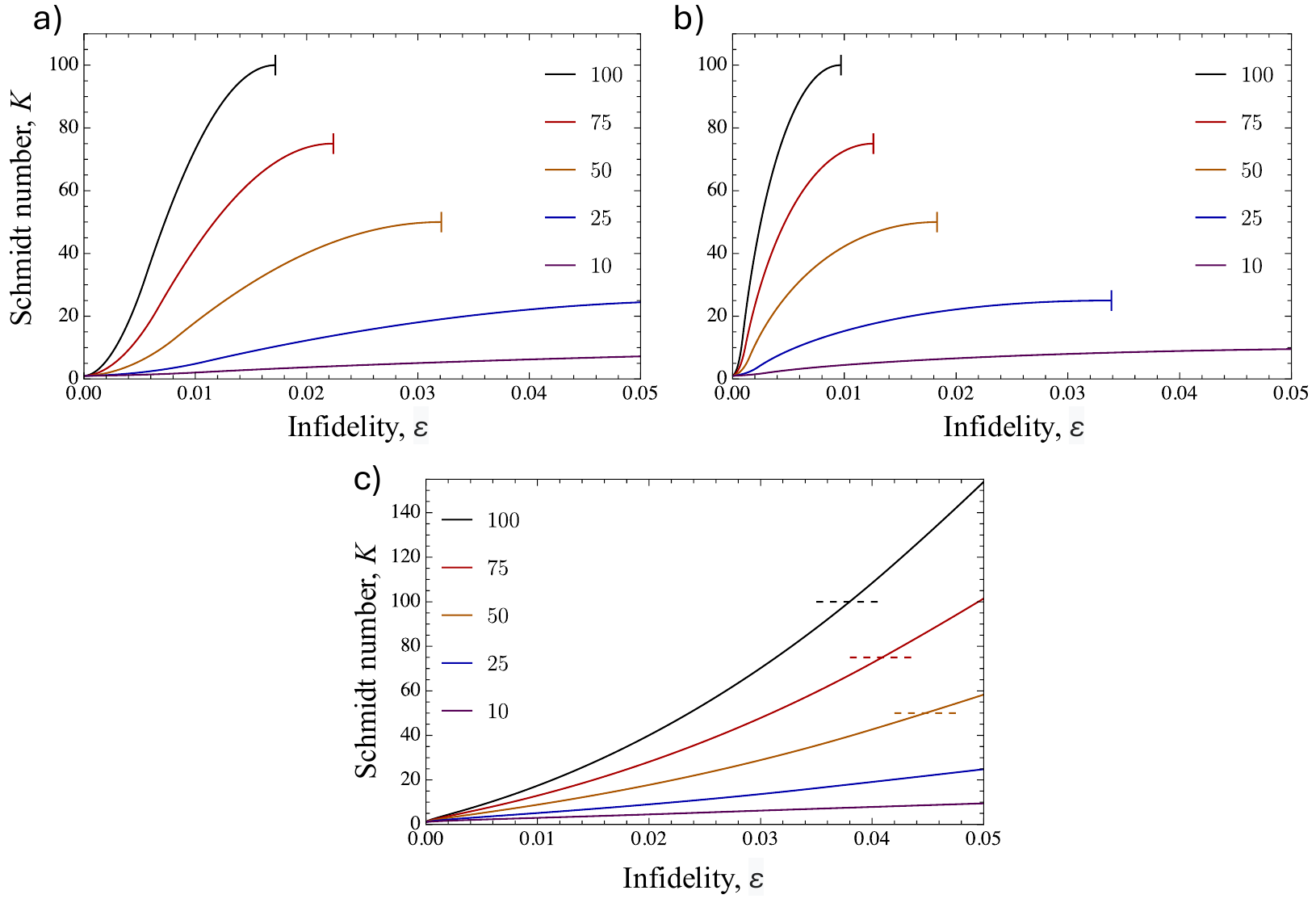}
    }
    \captionsetup{width=1\linewidth}
    \caption{False Schmidt number as a function of infidelity $\varepsilon$ for fixed dimension $d \in \{10,25,50,75,100\}$ for each of the three attack protocols: a) average fidelity, b) worst-case fidelity, and c) worst-case fidelity and individual projectors. The critical dimension at which the faked Schmidt number reach $K = d$ is marked with solid lines in a) and b), and with dashed lines in c).}
\label{fig:CompModelDim}
\end{figure}

\subsection{Impact of attacks on entanglement-to-separable gap} \label{sec:enttosep}
We have seen that at a given infidelity, the witness parameter $\mathcal{W}_{\text{prod}}^{(d)}(\varepsilon)$ increases with dimension. However, in the faithful entanglement witness test $\mathcal W_d$ - with perfectly implemented measurements - increasing the dimension is associated with a larger entanglement-to-separable gap. That is, the optimal violation of the witness remains constant, $\mathcal{W}_{\text{ent}}^{(d)} = 2$, whereas the separable bound, $\mathcal{W}_{\text{sep}}^{(d)} = 1 + 1/d$, decreases with increasing dimension. Therefore, we analyze how a small infidelity $\varepsilon$ impacts the witness parameter $\mathcal{W}_{\text{prod}}^{(d)}(\varepsilon)$ when increasing the dimension $d$. Since the standard entanglement criterion $\mathcal W_d$ is based on two MUBs, the effect of increasing $d$ has two competing effects: (i) the number of measurement elements of each MUB increases, and (ii) the distance between any two elements in the complementary bases increases. The former effect ought to increase the impact of $\varepsilon$, whereas the latter ought to decrease it. To study the trade-off between these effects, we compute ratio between the entanglement-to-separable gap in the adversarial and ideal case\cite{Morelli_2022}. To this end, we obtain that
\begin{equation}\label{eq:entsepgap}
\Delta \equiv \frac{\mathcal{W}_{\text{ent}}^{(d)}-\mathcal{W}_{\text{prod}}^{(d)}(\varepsilon)}{\mathcal{W}_{\text{ent}}^{(d)}-\mathcal{W}_{\text{sep}}^{(d)}}= \frac{d}{d-1}\left(2-\mathcal{W}_{\text{prod}}^{(d)}(\varepsilon)\right).
\end{equation}

This can be computed using the analysis of our three attack protocols. In Fig.~\ref{fig:Delta} we plot the entanglement-to-separable parameter $\Delta$ as a function of $d$ for fixed infidelity $\varepsilon$. We observe that for small dimensions the second effect is dominating; the detrimental impact of the adversarial deviations is remedied, but for larger dimensions the first effect takes over and $\Delta$ decreases steadily under the influence of the measurement deviations.

\smallskip

The analysis of the quantity $\Delta$ was first proposed by Morelli et.al.\cite{Morelli_2022}. However, there it was based only on brute-force numerical search which could not exceed $d=10$, thereby not accessing the most relevant regime for high-dimensional quantum information. As a consequence, the main effect for large dimensions did not clearly appear in the analysis. In our approach, with at most a single-parameter optimization, we can access arbitrary high dimensions and reveal the main behavior in the high-dimensional regime.

\begin{figure}[ht!]
    \makebox[\linewidth][c]{%
        \includegraphics[width=1\linewidth]{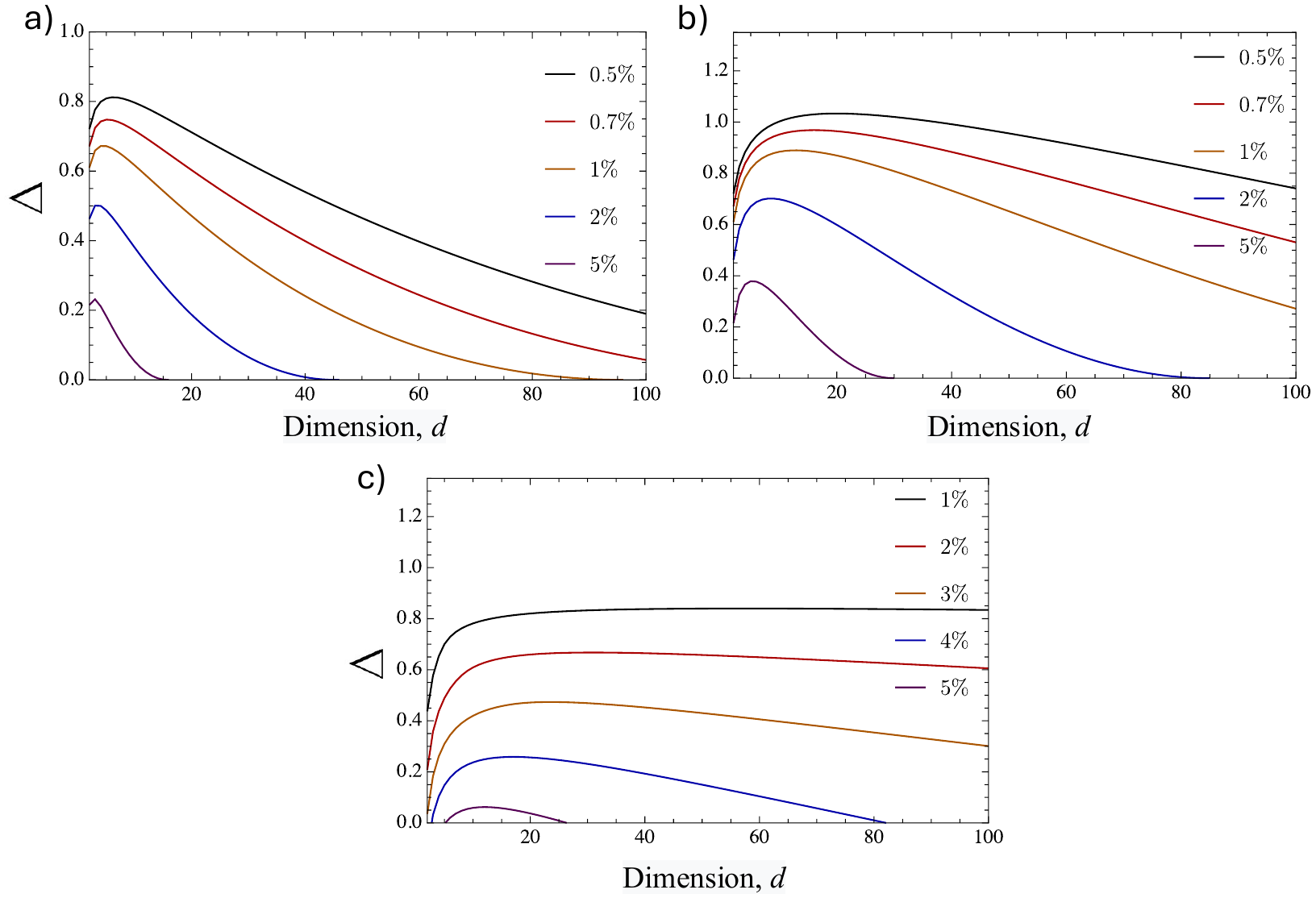}
    }
    \captionsetup{width=1\linewidth}
    \caption{The entanglement-to-separable gap $\Delta$ for fixed infidelities $\varepsilon$ for each of the three attack protocols: a) average fidelity, b) worst-case fidelity, c) worst-case fidelity and individual projectors. In a) and b) the infidelity is fixed to $\varepsilon \in \{0.5\%,0.7\%,1\%,2\%,5\%\}$, whereas in c) it is fixed to $\varepsilon \in \{1\%,2\%,3\%,4\%,5\%\}$.}
\label{fig:Delta}
\end{figure}

\subsection{Simulating probability for maximal entanglement}
We have seen that the optimal value of the original entanglement test $\mathcal W_d$, associated with maximal entanglement, can be achieved by using a product state and measurements that compromise a small adversarial deviation. However, a natural question is whether the attack protocols are able to not only fake entanglement, but also reproduce the full set of probabilities observed in the original experiment. Here, we show that this is possible by leveraging classical correlations.

In the original entanglement test $\mathcal W_d$, the probability that Alice and Bob obtain identical outcome, given that the parties share a maximally entangled state $\ket{\phi^+} = \frac{1}{\sqrt{d}}\sum_{i} \ket{i,i}$, is given by
\begin{equation}\label{eq:prob_me}
    p^{\text{targ}}(j,j) = \frac{1}{d}, \qquad p^{\text{targ}}(f_j,f_j^*) = \frac{1}{d}.
\end{equation}
Here, we have denoted $p^{\text{targ}}(j,j) = \langle j,j |\phi^+|j,j\rangle$ and $p^{\text{targ}}(f_j,f_j^*) =\langle f_j,f_j^*| \phi^+|f_j,f_j^*\rangle$, with $j = 0,\dots,d-1$. Note that each pair of identical measurement outcomes occur with uniform probability.

We aim to simulate the measurement statistics in Eq.~\eqref{eq:prob_me} using product states. To this end, consider that the source between Alice and Bob randomly and uniformly receives an input $\lambda \in\{0,\dots,d-1\}$ and encodes the symbol into a classical state of the form
\begin{equation}
\ket{\Psi^{(\lambda)}}_{AA'BB'}=\ket{\psi^{(\lambda)}}_A\otimes \ket{\psi^{(\lambda)}}_B \otimes \ket{\lambda}_{A'} \otimes \ket{\lambda}_{B'}.
\end{equation}
The systems $AA'$ are sent to Alice, whereas the systems $BB'$ are sent to Bob, where $A'$ and $B'$ are ancillary systems initialized in the computational basis state $\ket{\lambda}$. Upon receiving the system $AA'$, Alice's measurement apparatus first reads the classical register $\ket{\lambda}_{A'}$ and then uses the read-out to select a measurement to be performed on the state $\ket{\psi^{(\lambda)}}_{A}$. A similar procedure holds for Bob's measurement apparatus. Again, Alice's and Bob's lab measurements are optimally chosen to be equal. We label these measurements by $\tilde E_{j|\lambda}$ and $\tilde F_{j|\lambda}$, where the former and the latter are associated with the experimenters attempt at realizing the computational basis and the Fourier basis, respectively. For each $\lambda$, the measurement $\{\tilde E_{j|\lambda}\}_j$ and $\{\tilde F_{j|\lambda}\}_j$ must be $\varepsilon$-close to the target bases. The average probability that Alice and Bob obtain the same measurement outcome now reads
\begin{equation}\label{eq:prob_prod}
    p(j,j) = \frac{1}{d} \sum_\lambda \langle \psi^{(\lambda)} |\tilde E_{j|\lambda} |\psi^{(\lambda)}\rangle^2, \qquad p(f_j,f_j^*) = \frac{1}{d}\sum_\lambda \langle \psi^{(\lambda)} |\tilde F_{j|\lambda} |\psi^{(\lambda)}\rangle^2.
\end{equation}
To quantify how close the probability distribution in Eq.~\eqref{eq:prob_prod} is to the target distribution in Eq.~\eqref{eq:prob_me}, we use the statistical distance measure. For example, for distributions associated with the computational basis this measure reads
\begin{equation}
    \Xi(p^{\text{targ}},p) = \frac{1}{2}\sum_{j =0}^{d-1} |p^{\text{targ}}(j,j) - p(j,j)|,
\end{equation}
and similarly holds for the Fourier basis. 

We now apply the described procedure to the attack protocols based on average fidelity and worst-case fidelity. However, the same procedure cannot straightforwardly be applied to the third attack. The reason is that the set of individual projections do not have to form a valid measurement, which results in the associated measurement statistics not forming a valid probability distribution.

\subsubsection{Attack for average fidelity}\label{sec:avfid_classicalrandom}
We first study the measurement statistics generated by the attack protocol based on average fidelity by leveraging classical correlations. To each variable $\lambda$, we associate a product state $\ket{\psi^{(\lambda)}} = U^\lambda \ket{\psi}$ and lab measurements
\begin{equation}
\begin{aligned}
    \tilde E_{k|\lambda} = U^\lambda \ketbra{\phi_{k-\lambda}}{\phi_{k-\lambda}} (U^{\lambda}){}^\dagger, \qquad \tilde F_{k|\lambda} = U^\lambda G \ketbra{\phi_{k-\lambda}}{\phi_{k-\lambda}} G^\dagger (U^{\lambda}){}^\dagger,
    \end{aligned}
\end{equation}
where $\ket{\psi}$ is the optimal product state defined in Eq.~\eqref{stateansatz} and $\{\ket{\phi_k}\}_k$ is the optimal basis defined in Eq.~\eqref{eq:meas_ave}. We further define the unitary operator as follows $U^\lambda = X^\lambda Z^\lambda$, where the so-called shift-and-clock operators are given by $X = \sum_{j=0}^{d-1} \ketbra{j+1}{j}$ and $Z = \sum_{j=0}^{d-1} \omega^j \ketbra{j}{j}$, respectively, with $\omega = e^{2 \pi i/d}$. For each $\lambda$, the overlaps between the perturbed computational basis and its target basis are given by 
\begin{equation}
    \braket{k|\tilde E_{k|\lambda}|k} = \lvert\langle k|U^\lambda|\phi_{k-\lambda}\rangle \rvert^2 = \lvert\braket{k-\lambda| Z^\lambda|\phi_{k-\lambda}}\rvert^2 = \lvert \braket{k-\lambda|\phi_{k-\lambda}}\rvert^2.
\end{equation}
Similarly, for the Fourier basis we find that
\begin{equation}
\begin{aligned}
\braket{f_k| \tilde F_{k|\lambda} |f_k}=\lvert \langle k| G^\dagger U^\lambda G |\phi_{k-\lambda}\rangle \rvert^2 = \lvert \braket{k-\lambda|G^\dagger X^\lambda G |\phi_{k-\lambda}}\rvert^2 = \lvert\braket{k-\lambda|\phi_{k-\lambda}}\rvert^2.
    \end{aligned}
\end{equation}
Here, we have used the relations $G X = Z G$ and $ZX = \omega X Z$. Thus, the average fidelity constraint is fulfilled for each $\lambda$ by construction, $\frac{1}{d} \sum_k \braket{k|\tilde E_{k|\lambda}|k} = \frac{1}{d} \sum_k \lvert\braket{k-\lambda|\phi_{k-\lambda}}\rvert^2 \geq 1-\varepsilon$ and $\frac{1}{d}\sum_k\braket{f_k| \tilde F_{k|\lambda} |f_k} =\frac{1}{d}\sum_k \lvert\braket{k-\lambda|\phi_{k-\lambda}}\rvert^2 \geq 1-\varepsilon$. Next, we compute the overlaps between the product state and the measurements. In the first basis the overlaps are
\begin{equation}
    \braket{\psi^{(\lambda)}|\tilde E_{k|\lambda}|\psi^{(\lambda)}}= \begin{cases}
        &\lvert\braket{\psi|\phi_{0}}\rvert^2,  \qquad k=\lambda \\
        & \lvert\braket{\psi|\phi_{k-\lambda}}\rvert^2, \quad k\neq\lambda.
    \end{cases} 
\end{equation}
In the second basis the overlaps are
\begin{equation}
\braket{\psi^{(\lambda)}|\tilde F_{k|\lambda}|\psi^{(\lambda)}} =
    \begin{cases}
        &\lvert\braket{\psi|G|\phi_{0}}\rvert^2, \qquad k = \lambda \\
        &\lvert\braket{\psi|G|\phi_{k-\lambda}}\rvert^2, \quad k \neq \lambda.
    \end{cases}
\end{equation}
The overlaps between the state and basis measurements are given by Eqs.~\eqref{eq:overlaps_ave_comp}-\eqref{eq:overlaps_ave_four}. By stochastically coordinating measurement strategies, we find that the average probability that Alice and Bob obtain identical outcomes reads
\begin{equation}
\begin{aligned}
        &p(j,j) = \frac{1}{d}\left(\lvert \langle \psi|\phi_0\rangle\rvert^4 + (d-1)\lvert \langle \psi|\phi_1\rangle\rvert^4 \right), \\
&        p(f_j,f_j^*) = \frac{1}{d}\left(\lvert \langle \psi|G|\phi_0\rangle\rvert^4 + (d-1)\lvert \langle \psi|G|\phi_1\rangle\rvert^4 \right),
\end{aligned}
\end{equation}
for all $j \in \{0,\dots,d-1\}$. This strategy results in the statistical distance between the target distribution and the lab distribution being
\begin{equation}
\begin{aligned}
       & \Xi^{comp}= \frac{1}{2}\big\lvert 1-\lvert \langle \psi|\phi_0\rangle\rvert^4 - (d-1)\lvert \langle \psi|\phi_1\rangle\rvert^4 \big\rvert\\   
       &  \Xi^{four}= \frac{1}{2}\big\lvert 1 - \langle \psi|G|\phi_0\rangle\rvert^4 - (d-1)\lvert \langle \psi|G|\phi_1\rangle\rvert^4\big\rvert.
\end{aligned}
\end{equation}
Note that these expressions measure the statistical distance and are similar to the entanglement-to-separable gap $\Delta$. 

In Fig.~\ref{fig:Xi} we plot the statistical distance $\Xi^{comp}$ between the probability distribution associated with maximal entanglement and lab measurements as a function of $d$ for fixed infidelity $\varepsilon$. We observe that for small dimensions the distance between the target distribution and the lab distribution increases, but for larger dimensions $\Xi^{comp}$ steadily decreases under the influence of measurement deviations. One can easily confirm that $\Xi^{four}$ gives the same result.

\subsubsection{Attack for worst-case fidelity}
Next we study the attack protocol based on the worst-case fidelity, following the same procedure as presented above. That is, to each $\lambda$, we associate a product state $\ket{\psi^{(\lambda)}} = U^\lambda \ket{\psi}$ and lab measurements
\begin{equation}
\begin{aligned}
    \tilde E_{k|\lambda} = U^\lambda E_{k-\lambda} (U^{\lambda}){}^\dagger, \qquad \tilde F_{k|\lambda} = U^\lambda G E_{k-\lambda} G^\dagger (U^{\lambda}){}^\dagger,
    \end{aligned}
\end{equation}
where $\ket{\psi}$ is defined in Eq.~\eqref{stateansatz} and the POVM $\{E_k\}_k$ is given in Eq.~\eqref{eq:meas_worst}. The worst-case fidelity constraint is satisfied for each $\lambda$ by construction, since $ \braket{k|\tilde E_{k|\lambda}|k} = \braket{k-\lambda|E_{k-\lambda}|k-\lambda} \geq 1-\varepsilon$, and $\braket{f_k| \tilde F_{k|\lambda} |f_k} = \braket{k| G^\dagger \tilde F_{k|\lambda} G|k} = \braket{k-\lambda|E_{k-\lambda}|k-\lambda} \geq 1-\varepsilon$. By computing the overlaps between $\ket{\psi^{(\lambda)}}$ and the elements of the first POVM we find that $\langle \psi^{(\lambda)}|\tilde E_{k|\lambda}|\psi^{(\lambda)}\rangle = \langle \psi|E_{k-\lambda}|\psi\rangle$, where
\begin{equation}
\langle \psi|E_{k-\lambda}|\psi\rangle =
\begin{cases}
&\lvert \braket{\psi|\phi_{0}^\uparrow}\rvert^2 + \lambda_1  \lvert\braket{\psi|\phi_{0}^\downarrow}\rvert^2, \qquad k=\lambda \\
& \lambda_2\lvert\braket{\psi|\phi_{k-\lambda}} \rvert^2, \qquad \qquad \quad \ \ \ k\neq \lambda.
\end{cases}
\end{equation}
The overlaps $\braket{\psi|\phi_k}$ are given in Eq.~\eqref{eq:wc_overlap_comp}. Similarly, the overlaps between $\ket{\psi^{(\lambda)}}$ and Fourier transformed elements read $\langle \psi^{(\lambda)}|\tilde F_{k|\lambda}|\psi^{(\lambda)}\rangle = \langle \psi|G E_{k-\lambda} G^\dagger|\psi\rangle$, where 
\begin{equation}
\langle \psi|G E_{k-\lambda} G^\dagger|\psi\rangle =
\begin{cases}
&\lvert \braket{\psi|G|\phi_{0}^\uparrow}\rvert^2 + \lambda_1  \lvert\braket{\psi|G|\phi_{0}^\downarrow}\rvert^2, \qquad k=\lambda \\
& \lambda_2 \lvert\braket{\psi|G|\phi_{k-\lambda}} \rvert^2, \qquad \qquad \qquad \quad k\neq \lambda.
\end{cases}
\end{equation}
The overlaps $\braket{\psi|G|\phi_k}$ are given in Eq.~\eqref{eq:wc_overlap_four}. Since the probability that the parties obtain identical outcomes $p(j,j)$ and $p(f_j,f_j^*)$ is equal for all $j \in \{0,\dots,d-1\}$, the statistical distance between the target distribution and the lab distribution reads
\begin{equation}
\begin{aligned}
       & \Xi^{comp}= \frac{1}{2}\bigg\lvert 1-\left(\lvert \braket{\psi|\phi_{0}^\uparrow}\rvert^2 + \lambda_1  \lvert\braket{\psi|\phi_{0}^\downarrow}\rvert^2\right)^2 +(d-1)\lambda_2^2\lvert\braket{\psi|\phi_{1}} \rvert^4 \bigg\rvert,\\   
       &  \Xi^{four}= \frac{1}{2}\bigg\lvert 1-\left(\lvert \braket{\psi|G|\phi_{0}^\uparrow}\rvert^2 + \lambda_1  \lvert\braket{\psi|G|\phi_{0}^\downarrow}\rvert^2\right)^2 +(d-1)\lambda_2^2\lvert\braket{\psi|G|\phi_{1}} \rvert^4 \bigg\rvert.
\end{aligned}
\end{equation}
In Fig.~\ref{fig:Xi} we plot the statistical distance $\Xi^{comp}$ between the probability distribution associated with maximal entanglement and lab measurements when both parties attempt to measure in the computational basis. The qualitative behavior of the attack bases on worst-case fidelity is the same as for the attack based on average fidelity. That is, for small dimensions the distance between the target distribution and the lab distribution increases, but for larger dimensions it steadily decreases. Moreover, one can verify that $\Xi^{four}$ gives the same result.

\begin{figure}[ht!]
    \makebox[\linewidth][c]{%
        \includegraphics[width=1\linewidth]{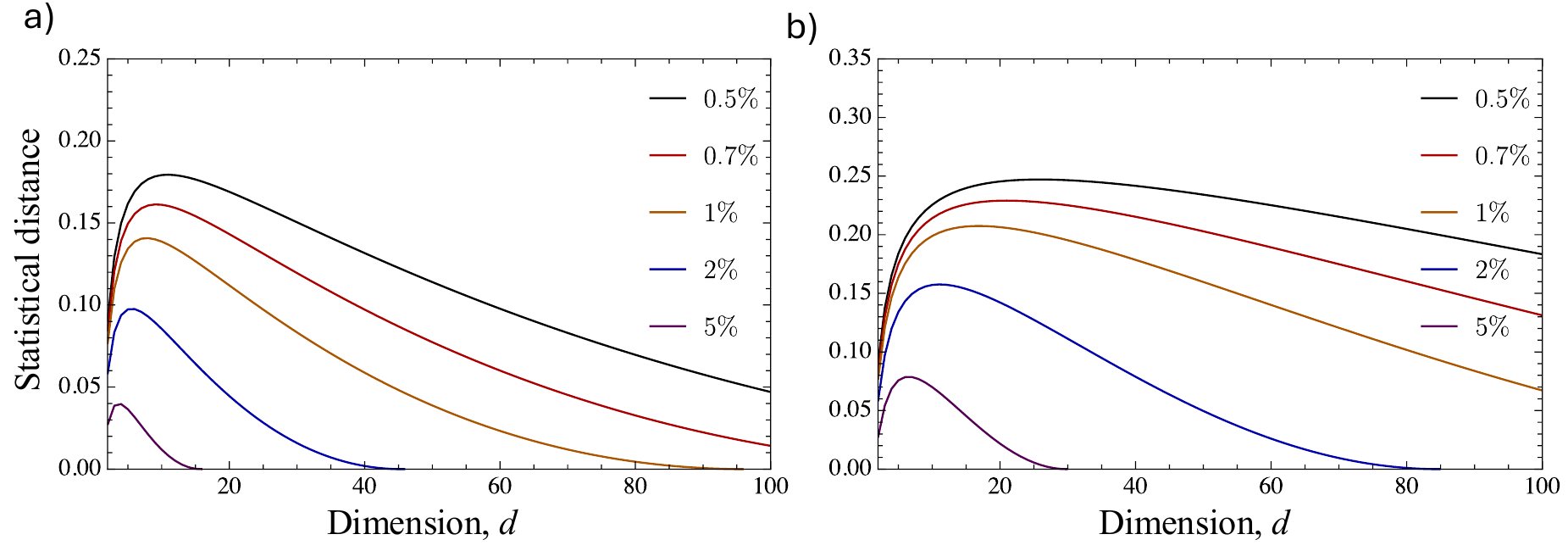}
    }
    \captionsetup{width=1\linewidth}
    \caption{The statistical distance $\Xi^{comp}$ between the target distribution associated with maximal entanglement and the probability distribution generated by the attack protocol based on average fidelity a) and worst-case fidelity b) using classical correlations, with fixed infidelity $\varepsilon \in \{0.5\%,0.7\%,1\%,2\%,5\%\}$.}
\label{fig:Xi}
\end{figure}

\section{Experimental details}
\subsection{ Method for state generation}\label{sec:Suppexp_generation}

Our goal is to experimentally verify that a two-photon product state $\ket{\psi} = \ket{\psi}_A\otimes\ket{\psi}_B $ can surpass the bounds in Eq.~\eqref{eq:SN}. 
The shared state $\ket{\psi}_{A,B}$ between Alice and Bob (indicated by the indices $A$ and $B$) can be generated with an attenuated laser as no quantum correlations are required. The only experimental requisite is to have a time reference between two photons for efficient coincidence counting in the measurement stage.
This timing correlation is achieved by using an attenuated laser pulse. The laser pulse is generated by frequency doubling a femto-second laser with a wavelength of $1560 \,$nm (Origami, NKT photonics) and a pulse duration of $220 \,$fs.
The up-converted pulse with a wavelength of $780 \,$nm is sent through a $5 \,$m single mode fiber (SMF) to the setup shown in Fig.~\ref{fig:ExpSetup}\,a). The estimated pulse duration is $\sim 2.5 \, $ps due to dispersion during the propagation along the fiber, which is still below the timing resolution of our single photon detectors of around  $\sim 350 \, $ps. 
The laser polarization is adjusted using wave-plates to match the required polarization for best operation of the spatial light modulators (SLMs). The laser is attenuated to a few-photons per pulse using a variable neutral density (ND) filter. For simplicity, from now on we will use ``two-photon'' and ``split attenuated pulse laser'' interchangeably.

\begin{figure}[h!]
    \centering
    \includegraphics[width=1\linewidth]{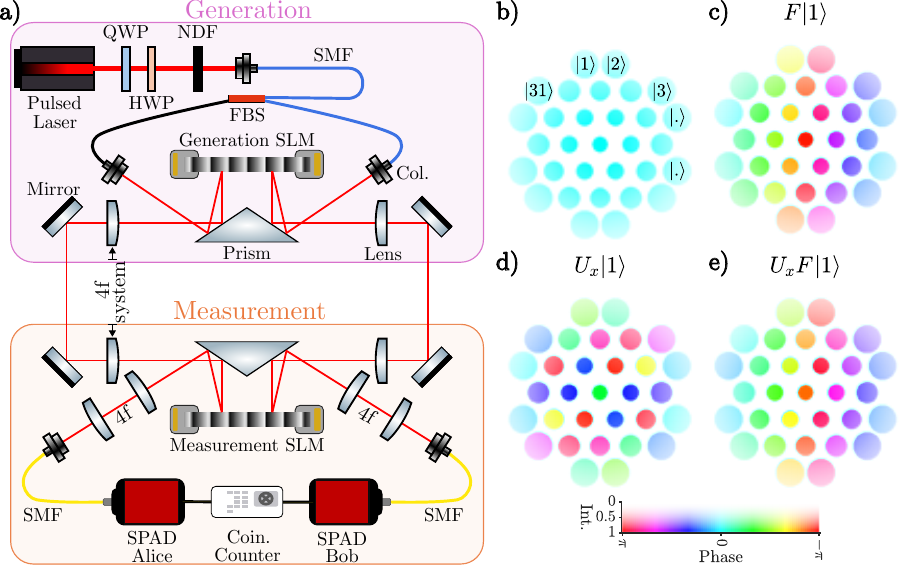}
    \caption{Schematic of the experimental scheme and macro-pixel basis implemented with the SLMs. a) The generation stage can produces any two-photon product state in the macro-pixel basis, while the measurement stage performs arbitrary local projections on each photon. b) Representation of all the qudit states $\ket{i}$ in the computational basis for $d = 31$. c) Corresponding first state $\ket{1}$ of the Fourier basis. d) First state of the computational basis $U_x\ket{1}$ of the balanced macro-pixel basis. e) Corresponding first state $U_x\ket{1}$ of the Fourier basis. QWP: Quarter wave plate, HWP: Half wave plate, NDF: Neutral density filter, FBS: Fiber beam splitter, Col: Collimator, SLM:Spatial light modulator, SMF: Single mode fiber, SPAD: Single photon avalanche diode.}
    \label{fig:ExpSetup}
\end{figure}

The transverse profile of a light field forms a continuous space. 
It is standard to describe this space using an infinite dimensional mode basis, such as Laguerre-Gauss modes and then restrict the space to a given mode number to effectively realize a finite-dimensional space \cite{Forbes2025}. In our case, we choose to discretize the transverse modes using the macro-pixel basis instead of Laguerre-Gauss modes (or another propagation-invariant mode family) as it provides better versatility and scalability to high dimensions \cite{valencia2020high}.
 
The attenuated laser is split into two paths using a $50\!:\!50$ fiber beam splitter (FBS). The desired spatial mode in each path is carved from the initial gaussian profile using the generation SLM and a mode carving technique, which allows for simultaneous phase and amplitude manipulation \cite{Bolduc:13}.
The displays of the generation and measurement SLMs are split into two to accommodate Alice's and Bob's paths with only two SLMs instead of four.
The two pulses are then relayed to the measurement stage using a $4f$ system with two lenses of equal focal length.
The lens system is required because the macro-pixel states are not eigenmodes of free space, i.e., they are not invariant under propagation, which means that an imaging sysyem is required to keep the states in the same macro-pixel basis between the generation and measurement SLMs. 
Fig.~\ref{fig:ExpSetup}\,b) shows all the states for the macro-pixel basis in dimension $d=31$, each disk represents a state $\ket{i}$ of the computational basis. 
 
To properly generate a basis it is required that the photon flux $\Phi_i = N_i/\Delta t$ is the same among all macro-pixel disks in Fig.~\ref{fig:ExpSetup}\,b). To account for the transverse Gaussian intensity profile of the beam from which the pixel basis is carved, the radius $r$ of each disk is calculated depending on their position relative to the center of the input Gaussian profile. This is achieved by minimizing $r$ in the following formula 

\begin{equation}\label{eq:diskrad}
    \frac{1}{4}\left[ Erf \left(\frac{r-R}{\omega_0 } \right) +  Erf \left(\frac{r+R}{\omega_0 } \right)  \right]^2 = \Phi,
\end{equation}

here, $R = \sqrt{x^2+y^2}$ is the position of the disk center, $Erf()$ is the Gaussian error function and $\omega_0$ the beam waist radius of the input light field at the SLM plane. In our case the waist radius is $1.63 \pm 0.02 \,$mm. After the flux $\Phi_i$ is fixed for all disks, any superposition state can be properly generated from the basis states $\ket{i}$. 
Fig.~\ref{fig:ExpSetup}\,c) shows the first state of the Fourier basis $F\ket{1}$.
As outlined in the Methods section in the main text, the intensity mismatch between states in the computational and Fourier bases complicates the experiment, because the resulting count rates differ greatly between the two bases, with their intensities varying by a factor of $d^2$.
The SLM could be used to control the intensity of the states in the Fourier basis, but this would require to reduce the efficiency of the diffraction gratings by a factor of $d$, which significantly reduces the modulation quality of the SLM due its limited resolution in phase modulation steps (8 bit phase depths). 
A more effective approach is to introduce a unitary transformation $U_x$, as described in the Methods section, so that both the computational and Fourier basis states consist of $d$-disks and consequently have equal intensities. The first state of the transformed computational and Fourier basis are shown in Fig.~\ref{fig:ExpSetup}\,d) and e). The disk in the transformed states are now equally populated in both basis.

\subsection{Method for state detection}\label{secSM:detections}
The measurements in Eq.~(\ref{eq:SN}) rely on implementing projection operators of the form $E_j^A \otimes E_j^B \ketbra{\psi_{AB}}{\psi_{AB}}$. From the spatial mode picture, we need to construct a spatial mode filter for the macro-pixel basis. This can be achieved by using a phase/intensity flattening technique with an SLM and SMF\cite{Bouchard_2018}. 
After coupling into the SMF, the photons are detected by single photon avalanche photo diodes (SPAD, Laser Components COUNT T series) with an approximate detection efficiency of $\sim 60-70 \, $\% at the operating wavelength and a timing jitter of around $\sim 350 \, $ps.
The detection signals are recorded by a time tagger (IDQuantique ID900) and coincidence detections were counted using a coincidence window of $2 \, $ns.
Given the symmetry of the setup (see Fig. \ref{fig:ExpSetup}\,a) ), the detection method for one path works equally for the other. 
The photon flux $\Phi_{ij}$ after the SMF, i.e., after preparing the state $\ket{i}$ with the generation SLM and projecting on the state $\bra{j}$ with the measurement SLM, is given by the overlap integral

\begin{equation}\label{eq:modecouling}
    \Phi_{ij} = \eta \Phi_{in} \left|\int_{x y}G(\omega_0, x,y)P^{(g)}_{i}(x, y) \left [P^{(m)}_{j}(x,y)G(\omega_c)\right]^* dxdy\right|^2.
\end{equation}

Here, $\eta$ is the overall system loss which is equal for all states (such as detector and SLM efficiencies), $\Phi_{in}$ is the initial photon flux from the source after the FBS, $G(\omega_0)$ is the input Gaussian profile, $P^{(g)}_{i}$ and $P^{(m)}_{j}$ are the holograms displayed on the SLMs at the generation and measurement stages, respectively, and $G(\omega_c)$ is the Gaussian profile propagated back from the detection SMF.
The latter gaussian mode is normally labeled as collection mode. The relation between the measured photon flux $\Phi_{ij}$ and the detection probabilities in Eq.~(\ref{eq:SN}) is given by

\begin{equation}\label{eq:fluxtoprob}
    \Phi_{ij} = \Phi_T \tr\left(E_j \ketbra{i}{i}\right), 
\end{equation}
with 
\begin{equation}
    \Phi_T = \sum_{j=0}^{d-1} \Phi_{ij},
\end{equation}

being the total photon flux measured from the projection in all states of the given basis. 
\\

In Eq.~(\ref{eq:fluxtoprob}), we have disregarded mode-dependent losses, but our experiment is not free of them. 
There are two major sources of mode-dependent losses. 
One is optical aberrations introduced by SLMs and lenses.
The second is that the radius of each disk (as shown in Fig.~\ref{fig:ExpSetup}\,b) is designed for the input beam waist $\omega_0$, not the collection mode waist $\omega_c$.
This means that the disks after the SLM in the measurement stage are not equally collected by the SMF, hence introducing mode-dependent losses in the measurement system. 
To visualize this effect, let's use the original computational basis shown in Fig.~\ref{fig:ExpSetup}\,b) with each disk being a state.
Fig.~\ref{fig:Expampcorrection}\,a) shows the comparison between two cases, for $\omega_c=\omega_0$ and $\omega_c=2\omega_0$. When the beam waists match there is an almost equal detection probability, calculated with Eq.~(\ref{eq:fluxtoprob}), for all basis states.
However, when the waists do not match it is more likely to detect certain states. In our experiment, the detection beam waist $\omega_c$ is fixed by the focal length of the microscope objective.
To overcome this deficiency, we use the SLM to compensate for mode dependent losses. 

The natural approach would be to display a modified hologram in the measurement state, $$P^{(m)'}_j =  P^{(m)}_j \frac{G(\omega_0)}{G(\omega_c)},$$ effectively matching the beam waist between input and collection mode at the SLM plane. 
Although this approach works in theory, in practice this correction assumes that optical aberrations are negligible, which is not the case in most optical setups. Optical aberrations can be understood as phase distributions which, in theory, can be fully compensated using a SLM.
In our experiment, we use the Zernike polynomial scanning technique to compensate for aberrations \cite{AdaptiveOpticsBMJ2007} to a great extent.
However, this is not enough to achieve the required fidelities as the variations between the coupling of states are around $8\%$ for dimension $31$. A more suitable approach is to perform a custom correction for each state on the measurement SLM by multiplying each disk of the computational basis by a factor $\alpha_jP^{(m)}_j$.
To calculate $\alpha_j$, we measure the count rates $C_j = \Phi_j \Delta t$ for each single macro-pixel and define $\alpha_j = \sqrt{\frac{\text{min}(C)}{C_j}}$, such that the macro-pixel with the lowest count rate stays fix while the other decrease in amplitude.
Due to experimental fluctuations, the values of $\alpha_j$ varies, so we perform this measurement iteratively until $\text{var}(C)$ reaches an stable minimum. 
Using this procedure, we are effectively constructing an custom amplitude mask $ \alpha(x,y) = \sum_i \alpha_i$ as each $\alpha_i$ correspond to different spatial locations on the SLM.
This mask compensates for the different coupling efficiencies and works for any superposition states. 
Fig.~\ref{fig:Expampcorrection}\,b) shows the simulated detection probabilities of an arbitrary state $\ket{\psi_r}$ in the computational basis using the compensation methods: beam waist matching and individual macro-pixel amplitude.
The error between the simulated detection probabilities and theory is less than $0.5\%$ for several simulated states. 
Although both methods give similar results in simulations, we have found that the custom compensation performs better experimentally. 

Figs.~\ref{fig:Expampcorrection}\,c) and d) show the action of the custom compensation in our experiment for $d=31$ and $d=61$, respectively. For $d=31$, the initial variation in detection probability among all basis states is $20.3 \pm 0.2 \,\%$, which is reduced to $2.1 \pm 0.2 \,\%$ after compensation. 
In the case of $d=61$, the initial variation is $14.7\pm 0.2 \,\%$, which is reduced to $3.7 \pm 0.2 \,\%$. 
We note that we expect that the custom, individual macro-pixel compensation, starts to fail when the disk sizes are big enough 
such that the Gaussian profile of the collection mode cannot be approximated by a constant $\alpha_i$ inside the disk.
\\

\begin{figure}[h!]
    \centering
    \includegraphics[width=1\linewidth]{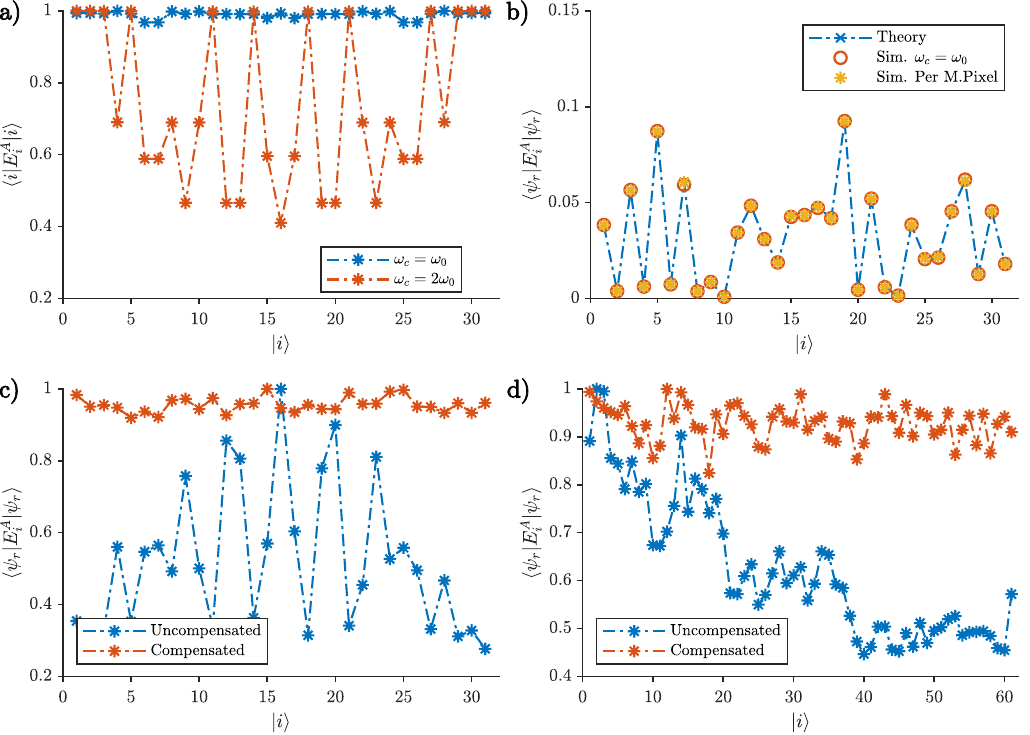}
    \caption{Amplitude correction for the collection modes. a) Difference between the coupling efficiency of each disk for two different ratios between input and collection mode waist. b) Simulation of the detection of a random state by compensating using the collection mode Gaussian profile $\omega_c = \omega_0$ or custom compensation (Per Macro-Pixel) using the mask $\alpha(x,y)$ . Measured efficiencies of all states when the custom compensation method is used for c) $d=31$ and d) $d=61$, respectively.}
    \label{fig:Expampcorrection}
\end{figure}

We now turn back to the implementation of the projective measurements. 
The joint projector $E_j^A \otimes E_j^B $ is in a factorized form, therefore each projector can be implemented locally by properly choosing the hologram $P^{(m)}_j$ before the SMF coupling to project the state $\bra{j}$ in each path. 
Considering that both state and projectors are factorizable, the measurements can be obtained from independent detections from the right and left path of the experiment.
However, our goal is to measure an entanglement witness without any prior knowledge of the input state, to this end we use the timing correlation of the initial attenuated pulse and coincidence counting to mimic the measurement of photon pairs. 
Given an arbitrary two-photon state $\ket{\psi_{AB}}$, the coincidence count rate [counts/sec] is given by 

\begin{equation}\label{eq:cointoprob}
    C_{ij} = C_T \tr\left(E_i^A \otimes E_j^B \ketbra{\psi_{AB}}{\psi_{AB}}\right),
\end{equation}
with 
\begin{equation}
    C_T = \sum_{i,j=0}^{d-1} C_{ij},
\end{equation}

being the total coincidence counts from all states in the two-photon basis. Therefore, to convert the diagonal coincidence count rates $C_{ii}$, required in Eq.~(\ref{eq:SN}), into detection probabilities, we must measure all $(ij)$ combinations of the projector $E_i^A \otimes E_j^B$ in order to determine $C_T$. In a multi-outcome measurement, all such combinations are inherently included, since each measurement yields $d$ possible outcomes. In our experiment, we can only carry out individual projections using the SLM.
However, by sequentially implementing all combinations of the individual projector $E_i^A \otimes E_j^B$, as it is common in most high-dimensional measurements in photonics, we can reproduce the statistics of a multi-outcome measurement, such as the ones in the the two multi-outcome models discussed in sections~\ref{sec:average_model} and \ref{sec:worst_case}. 
In section~\ref{sec:IndProj_Gauged}, we introduce another method to transform counts to detection probabilities using only the diagonal projectors $E_i^A \otimes E_i^B$ lifting the requirement of measuring $C_T$, this method is used for the worst-case attack with individual projector described in section~\ref{sec:Bin_worst_case}.

\subsection{Crosstalk benchmarks}

To be able to certify that a product state surpasses the bounds in Eq.~(\ref{eq:SN}) with the smallest perturbation possible, it is desirable to start with a near perfect system and then introduce the small $\varepsilon$-perturbations described in the three models in sections \ref{sec:average_model}, \ref{sec:worst_case}, and \ref{sec:Bin_worst_case}. The major experimental difficulty is to reduce the crosstalk between the modes, that is, the non-zero detection probability, $|\braket{i|j}|^2\neq0$, for modes that should be orthonormal. In practice, many experimental imperfections, including aberrations and misalignment, can be effectively described as crosstalk. In our setup, one of the primary systematic errors arises from the mode-carving technique itself.
Because this method relies on displaying diffraction gratings on the SLM, it introduces coupling between frequency and spatial degrees of freedom, causing different frequencies to experience slight spatial shifts in the diffraction orders. This effect produces significant crosstalk between states because each frequency component is displaced in the measurement fiber plane. To mitigate this, we employ a 3\,nm band-pass filter centered at 780\,nm to narrow the laser bandwidth. 
Additionally, we apply the same diffraction grating with opposite sign on the generation and measurement SLMs. In this way, the frequency-dependent spatial shift is canceled at the fiber plane.

In addition, the SLMs also introduce crosstalk from phase aberrations, since any pressure points on the screen and/or temperature fluctuations induce unwanted phase changes. 
These aberrations are more prominent at the edges of the SLM, where the screen is mounted to the frame. Furthermore, temperature variations make long-term compensation challenging. 
In our experiment, we have managed to use the same correction over one to two days, which is not enough to collect all measurements presented in this article such that between models, the compensation procedure had to be repeated.

As mentioned above, to correct for phase aberrations, we implement an algorithm that decomposes the aberrations into Zernike polynomials \cite{AdaptiveOpticsBMJ2007}. For each SLM, the coefficients of the Zernike terms are determined by maximizing the rotational symmetry of Orbital Angular Momentum (OAM) modes generated by the SLM at their respective focal plane. This procedure is done for the four modulations planes shown in Fig.~\ref{fig:ExpSetup}a) using a camera. 
A final fine-tuning of the optimization is performed by minimizing the coupling of the OAM modes into a SMF, since coupling to a SMF is more sensitive to residual phase aberrations than the previous procedure. 

\begin{figure}[h!]
    \centering
    \includegraphics[width=1\linewidth]{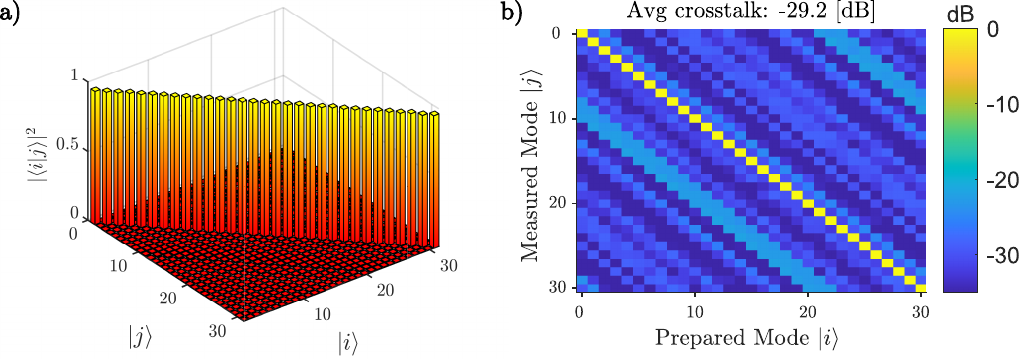}
    \caption{Detection probability crosstalk matrix for $d=31$ for a) modes measured in the computational basis, and b) for modes prepared in the computational basis and measured in the Fourier basis.}
    \label{fig:Expcrosstalk31_detP}
\end{figure}

The following measurements are performed using the basis presented in main text, that is, $ U_x\ket{j}$. 
Fig.~\ref{fig:Expcrosstalk31_detP}\,a) shows the detection probability for the mode $\ket{i}$, which is being displayed on the generation SLM while the mode $\bra{j}$ is displayed on the measurement SLM. The data corresponds to only one path of the setup, however, analogous results are obtained for the other path. 
The matrix shown in Fig.~\ref{fig:Expcrosstalk31_detP}\,a) is referred to as the crosstalk matrix and hereafter denoted as $S$.
It allows us to extract more informative quantifiers of the fidelity of our experimental implementation.
One quantifier is the average per-mode crosstalk $\bar{S}_D = \frac{1}{d}\sum_i|\braket{i|i}|^2 =  96.40  \pm 0.14 \,\% $ which implies an average detection probability of the wrong state of $\bar{S}_{XT}= \frac{1}{d(d-1)}\sum_{ij}|\braket{i|j}|^2 = 0.12\,\%$, $\forall j\neq i$.
The quantity $\braket{S_{off}}$ is commonly referred to as the crosstalk value. Another common approach to quantify crosstalk is to use detection ratios defined as 

\begin{align*}\label{eq:detRatio}
    D_{ij} &= \frac{\Phi_{ij}}{\Phi_{ii}},
\end{align*}

with $\Phi_{ij}$ being the count rates for a given measurement setting. For Fig.~\ref{fig:Expcrosstalk31_detP}\,a), the average detection ratio for all modes, that is, the average of all $D_{ij}$ for $i\neq j$, is $\braket{D_{i\neq j}} = 1/(834.1 \pm 1/16.2)$.
As these are small values, it is customary to report them in decibel scale [dB], using the relation $D_{ij}^{(dB)}=10 $log$_{10}(D_{ij})$, to have a better visualization of the data.
In this scale, it is easier to see that the crosstalk is not flat but it has a certain structure as can be seen in Fig.~\ref{fig:Expcrosstalk31_detP}\,b).
The structure depends on the overlap integral in Eq.~\eqref{eq:modecouling}. One can show that the overlap integral is the same when the index $\{i,j\}$ of the generation and measurement holograms $ \{P^{(g)}_i(x,y), (P^{(m)}_{j}(x,y))^*\}$ are shifted by the same amount.
This effect is illustrated by the diagonal lines in Fig.~\ref{fig:Expcrosstalk31_detP}\,b), given that the crosstalk between, for instance, mode $\ket{1}$ and $\ket{6}$ is similar to the crosstalk between $\ket{11}$ and $\ket{16}$.
Note that the achieved crosstalk values at the level of state-of-the-art implementations of high-dimensional systems using spatial modes of light \cite{HDgateLiu2026}. The difficulty of controlling a system scales dramatically with it's dimension.
Therefore, it is expected that our crosstalk values are higher than the two-qubit gate fidelities, reported to be around $99\%$ in state-of-the-art devices \cite{TransmonGatesFid2025, Evered2023}, 

Another benchmark is to measure how mutually unbiased the implemented computational and Fourier bases are.
From theory it is expected that $|\braket{i|F|j}|^2=1/d$, with $F$ being the discrete Fourier matrix. In the case of $d = 31$, we achieve a value of $1/d = 0.0323$. Fig.~\ref{fig:Expdim31_fourier} shows the detection probability between all state pairs. The average value of the detection probability is $0.032 \pm 0.004$ agreeing very well with the theory.
Nevertheless, if we calculate the fluctuations between all the pairs, the standard deviation is $0.0064$ which is around $20\%$ of the mean value.
This deviation shows that the system’s imperfections are small; however, some aberrations remain present even after aberration correction and compensation for mode-dependent losses.

\begin{figure}[h!]
    \centering
    \includegraphics[width=0.6\linewidth]{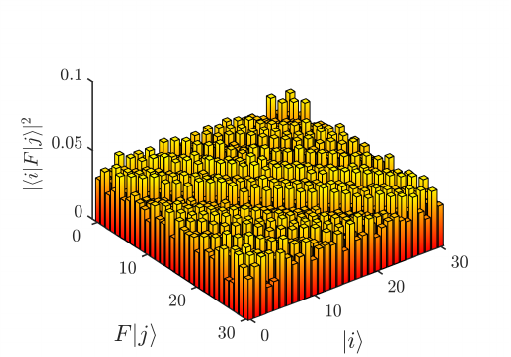}
    \caption{Detection probability of computational basis states measured in the Fourier basis.}
    \label{fig:Expdim31_fourier}
\end{figure}

The crosstalk matrices in dB for the computational basis in the other dimensions are show in Fig.~\ref{fig:Expct_otherdim}). The dimensions chosen for the experiments are 11, 19, 31, 43 and 61.
In theory there is no benefit in choosing prime dimensions given that the Fourier basis can be computed for any dimension.
However in practice, it is handy to be able to easily calculate any MUB as it helps with debugging imperfections in the system. The average detection ratio in dB for all dimensions is around $\braket{D_{i\neq j}} =-28.8 \pm 0.5 \,$dB which correspond to a detection ratio of $1\!:\!760$. To properly compare our measurement with theoretical predictions, we use a detection ratio of $1\!:\!700$ in theory as it fits better the experimental data. 
This discrepancy can be explained, as the crosstalk measurements are done on individual paths of the setup while the entanglement witness requires coincidence counting, hence the deviations of both paths accumulate. 
Moreover, the crosstalk increases with time due to temperature drifts and natural misalignment due to vibrations during the measurement time.

\begin{figure}[h!]
    \centering
    \includegraphics[width=1\linewidth]{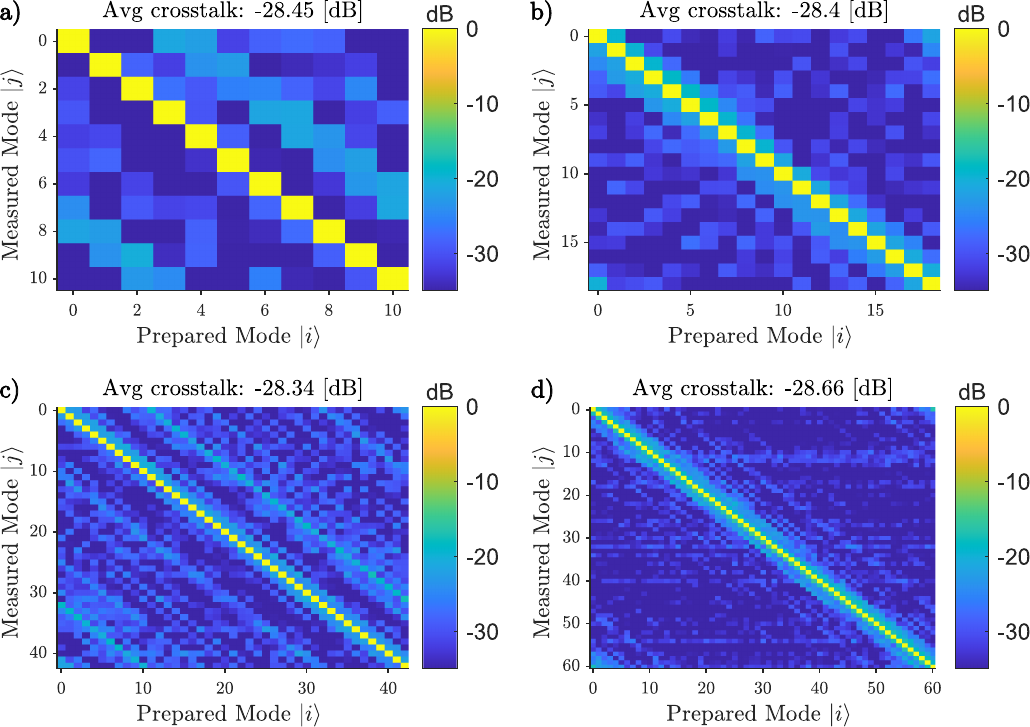}
    \caption{Crosstalk in dB for the computational basis with dimensions a) 11, b) 19, c) 43, and d) 61.}
    \label{fig:Expct_otherdim}
\end{figure}

\newpage
\subsection{Characterization of the imprecisions}

The imprecisions in both the average-fidelity model and the worst-case attack define a new POVM.
However to properly compare the experiment with the theory, we need to include our remaining experimental imperfection in the POVMs described in sections~\ref{sec:average_model} and \ref{sec:worst_case}. 
Here, our strategy is to benchmark the implemented imprecision by measuring the expectation values $\bra{i}E_j\ket{i}$ for each model and then inferring the implemented $\varepsilon$. In theory, we could have performed quantum process tomography to characterize our transformations, but the amount of required measurements scale exponentially with dimension, so it is not a reasonable choice for this experiment\cite{QPTChuang01111997}. 

The quantum measurement $\{\tilde{E}_l\}_{l}^{d}$ realized in the laboratory can be constructed from the ideal POVM $\{E_l\}_{l}^{d}$. 
Each element of the experimental POVM is defined as $\tilde{E}_i = \sum_{j=0}^{d-1} S_{ij} E_j$ as shown in the Methods section in the main text, where $S$ is referred to as the crosstalk matrix.
Using this relation one can compute the new bounds for the $\varepsilon$ perturbation for each model including the experimental crosstalk. 
In the main text, the subscript $\varepsilon_A$ was introduce to clarify the influence of the intrinsic experimental errors and deviations from an attack on the infidelity $\varepsilon$. For simplicity, we use $\varepsilon$ to refer only to the attacker deviation in the supplementary material.

\subsubsection{Average Fidelity model}
The fidelity bounds for the average model projector neglecting the crosstalks are given by

\begin{equation}\label{eq:ct_avemodel}
    \frac{1}{d}\sum_{j=0}^{d-1} \langle j|\tilde{E}_j|j\rangle = \bar{S}_{D}(1- \varepsilon),  
\end{equation}
where $\bar{S}_{D} = \frac{\text{Tr}(S)}{d}$ is the average detection probability. Then, one can estimate $\varepsilon$ and the crosstalk matrix $S$ from the measurements $\langle j|\tilde{E}_j|j\rangle$. 
The matrix $S$ correspond to the measurements shown in Fig.~\ref{fig:Expcrosstalk31_detP}\,a) for $d = 31$ and for the other dimension it can be extracted from the detection ratios in Fig.~\ref{fig:Expct_otherdim}.
The corresponding value of the average detection probability is $\bar{S}_{D} = 0.964 \pm 0.002$ for dimension $d=31$. It is worth mentioning, that the off-diagonal measurements $\langle i|\tilde{E}_j|i\rangle$ are still required to be measured to properly normalize the diagonal elements.

Fig.~\ref{fig:Channel_average} shows the measured expectation values for the imprecise projector $\langle i|\tilde{E}_j|i\rangle$ on all computational basis states $\ket{i}$ for $\varepsilon = 0.004$. Using Eq.~(\ref{eq:ct_avemodel}) we get an estimated $\varepsilon = 0.0044 \pm 0.0011$. Although the measurement agrees with the expected value, the uncertainty in the estimate is quite large because it depends on the measurement of $\text{Tr}(S)$. From Fig.~\ref{fig:Channel_average}\,a), one can observe that the imprecise basis is correctly implemented by the SLMs, since the expectation value $\langle 0|\tilde{E}_0|0\rangle$ is smaller than the other diagonal elements, which is in accordance with the theoretical prediction shown in Fig.~\ref{fig:Channel_average}\,b).

\begin{figure}[h!]
    \centering
    \includegraphics[width=1\linewidth]{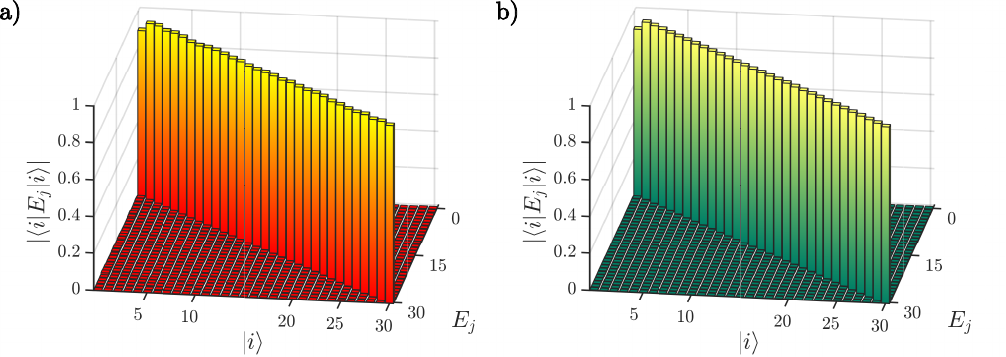}
    \caption{Average fidelity model. Expectation values of the computational basis states on the imprecise basis for $\varepsilon = 0.004$. a) Measurement. b) Theory assuming  average per-mode crosstalk $\bar{S}_{XT}= 0.115\%$}
    \label{fig:Channel_average}
\end{figure}

\subsubsection{Worst-case attack}
The fidelity bounds for the individual projectors in the worst-case attack with crosstalk are given by
\begin{equation}\label{eq:ct_inp}
     \langle j|\tilde{E}_j|j\rangle = \bar{S}_{D}(1- \varepsilon).  
\end{equation}
Fig.~\ref{fig:Channel_independent} shows the measurement and theoretical expectation values for $\varepsilon = 0.015$ in the worst-case attack model. The average value of $\braket{ j|\tilde{E}_j|j} = 0.9488 \pm 0.0012$ for all $j$, which gives an average $\braket{\epsilon} = 0.0158 \pm 0.0015$.

\begin{figure}[h!]
    \centering
    \includegraphics[width=1\linewidth]{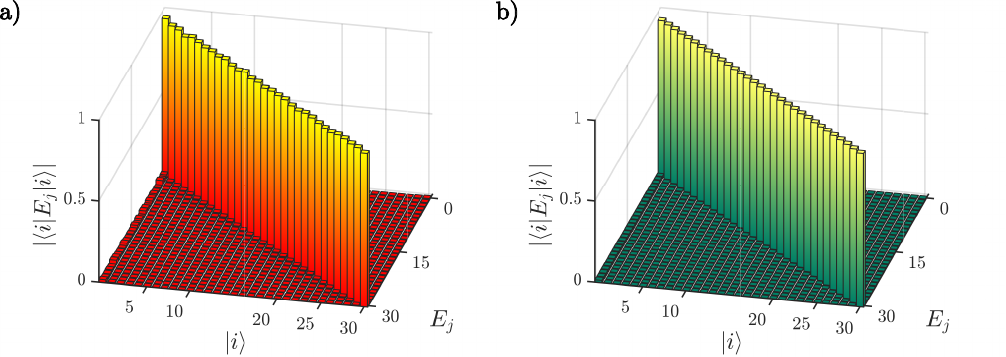}
    \caption{Worst-case fidelity model. Expectation values of the computational basis states on the imprecise basis for $\varepsilon = 0.015$. a) Measurement. b) Theory assuming an average per-mode crosstalk $\bar{S}_{XT}= 0.115\%$.}
    \label{fig:Channel_independent}
\end{figure}

\subsubsection{Worst-case attack with individual projectors}

In this model, the fidelity is bounded by Eq.~(\ref{eq:ct_inp}) as in the case of multi-outcome measurement. The difference is that the $\varepsilon$ perturbed measurements $\{E_j\}$ do not form a POVM, but each individual measurement is normalized as discussed in section~\ref{sec:Bin_worst_case}. Fig.~\ref{fig:Channel_binarized} shows the measured and theoretical expectation values for $\varepsilon = 0.015$. The average value of $\braket{ j|\tilde{E}_j|j} = 0.9491 \pm 0.0011$ for all $j$, which gives an average perturbation $\braket{\epsilon} = 0.0155 \pm 0.0014$.

\begin{figure}[h!]
    \centering
    \includegraphics[width=1\linewidth]{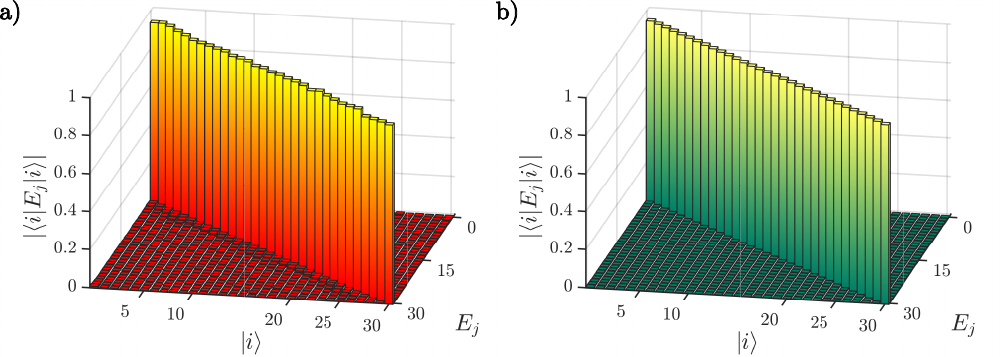}
    \caption{Worst-case fidelity model with individual projections. Expectation values of the computational basis states on the imprecise basis for $\varepsilon = 0.015$. a) Measurement. b) Theory assuming an  average per-mode crosstalk $\bar{S}_{XT}= 0.13\%$.}
    \label{fig:Channel_binarized}
\end{figure}

The results for each error model presented above indicate that the implementation of the $\varepsilon$ perturbations are faithful to the theory, despite our estimation only using the expectation values in the computational basis. A more robust benchmark of the imprecision would require quantum process tomography.
However, in our case, it suffices to show that the implemented perturbations are accurate enough to measure a Schmidt number $K > 1$ for a product state which indicates that we have managed to trick the device-user into measuring an incorrect entanglement witness, thus incorrectly classifying a product state as an entangled state.

\subsection{Individual projections with calibrated detection probabilities}\label{sec:IndProj_Gauged}
To transform counts or photon flux to detection probabilities using Eq.~(\ref{eq:fluxtoprob}), it is necessary to measure all states of one basis. This is the standard method in photonics, as it removes the requirement of characterizing the overall efficiency of your setup, although mode-dependent losses still need to be considered. However, alternative methods exist for converting count rates into detection probabilities.

Let's consider the POVM $\{E_j\}$ with each element a Rank-1 projector of the form $\ketbra{\phi_j}{\phi_j}$, the measured photon flux in Eq.~(\ref{eq:fluxtoprob}) after an individual projector $E_j$ given the input state $\ket{\nu}$ is

\begin{equation}\label{eq:fluxtoprobwitheff}
    \Phi_{\nu j} = \eta_\nu \eta_j\Phi_T \tr\left( \tilde{E}_j \ketbra{\nu}{\nu}\right),
\end{equation}

where $\tilde{E_j}$ denotes the experimental POVM—including crosstalk—derived in the Methods section in the main text. The quantities $\eta_{\nu}$ and $\eta_j$ represent the efficiencies associated with preparing the state $\ket{\nu}$ and implementing the projector $\tilde{E_j}$, respectively. 
As discussed above, both the states and the projectors are realized via amplitude/phase modulations on the SLM, which leads to mode-dependent efficiencies determined by the corresponding holograms and SMF coupling. The variations in the projector efficiencies $\eta_j$ are small because our POVM is constructed from slight perturbations of the computational/Fourier basis. Thus, we write $\eta_j = \eta \,\Delta\eta_j$, where $\eta$ is the average efficiency and $\Delta\eta_j \approx 1$ quantifies the small deviations between different POVM elements.

In the implementations discussed in section~\ref{secSM:detections}, the state efficiency $\eta_v$ and the average projector efficiency $\eta$ can be absorbed into the definition of $\Phi_T$. Then, the fluctuations $\Delta n_j$ effectively decrease the fidelity of the implemented POVM.
However, since these fluctuations are not under our control, we incorporate them directly into the description of the experimental POVM. With this treatment, Eq.~(\ref{eq:fluxtoprobwitheff}) simplifies to Eq.~(\ref{eq:fluxtoprob}), where all global efficiencies are absorbed by $\Phi_T$.

In the case of two-photon measurements, the equivalent quantity for $\Phi_T$ is the overall coincidence rate $C_T$ described in Eq.~(\ref{eq:cointoprob}). To estimate $C_T$ we would need to measure all combinations of the projector $E_i^A \otimes E_j^B$, even though not all of these measurement settings are required when classifying entanglement using Eq.~(\ref{eq:SN}). Our goal in this section is to lift the requirement of estimating $C_T$, such that we can compute the entanglement witness with only the diagonal projectors $E_i^A \otimes E_i^B$. 

Since our individual projectors are of the form $E_j = \ketbra{\phi_j}{\phi_j}$, we can express the photon flux via Eq.~(\ref{eq:fluxtoprobwitheff}) as

\begin{equation}
    \Phi_{jj} = \eta_j \eta\Phi_T \tr\left( \tilde{E}_j \ketbra{\phi_j}{\phi_j}\right) =\eta_j \eta\Phi_T S_{jj} \tr\left(E_j \ketbra{j}{j}\right) = \eta_j \eta\Phi_T S_{jj} ,
\end{equation}
where we separate the efficiencies from the total flux, since modifying the input state from $\ket{\nu}$ to $\ket{\phi_j}$ changes the generation efficiency. We can then compute the ratio between the photon flux for the two different input states,

\begin{equation}\label{eq:Bin_modlosses}
    \frac{\Phi_{\nu j}}{\Phi_{jj}} = \frac{\eta_\nu \tr\left( \tilde{E}_j \ketbra{\nu}{\nu}\right)}{ \eta_j S_{jj}}.
\end{equation}
Here, $\frac{\eta_\nu}{\eta_j}$ quantifies any mode-dependent losses in the generation of the states. In practice, the fluctuations of $S_{jj}$ between different measurements $E_j$ are small, so the diagonal crosstalk element $S_{jj}$ can be approximated by its average value $\bar{S}_{D}$ taken over all diagonal entries.

In practice, correcting for crosstalk during post-processing is not advised, since the crosstalk matrix changes over time due to mechanical drifts or temperature fluctuations. As a result, the matrix obtained from prior measurements may not accurately reflect the current state of the experimental setup, which would lead to unreliable compensation. Consequently, it is standard to assume the worst case scenario, i.e., there is zero knowledge about the crosstalk of the system, therefore $\bar{S}_{D} = 1$.

\begin{figure}[h!]
    \centering
    \includegraphics[width=0.6\linewidth]{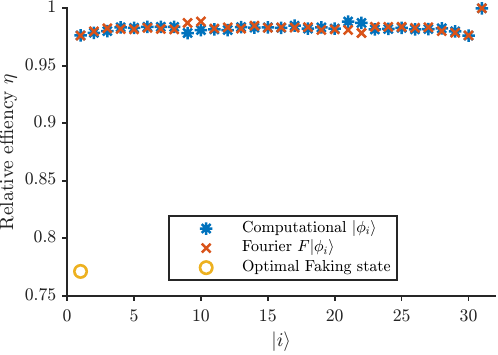}
    \caption{Relative efficiencies in the generation of the imprecise computational and Fourier bases states, and the optimal faking state for the average fidelity model with $\varepsilon = 0.007$}
    \label{fig:Modeefficiencies}
\end{figure}

Using Eq.~(\ref{eq:Bin_modlosses}) and knowing the photon flux for each state that generates the rank-1 projectors, the measured photon fluxes $\Phi_{\nu j}$ can be converted into detection probabilities. The only remaining step is to determine the ratios $\frac{\eta_\nu}{\eta_j}$ for all $j$. 
In our setup, the dominant contribution to mode-dependent losses arises from the mode carving in the SLM, as illustrated in Fig.~\ref{fig:Modeefficiencies}. 
In practice, to achieve the maximum efficiency for the desired hologram, the field $P^{(g)}(x,y)$ is typically normalized as 
\begin{equation}\label{eq:holonorm}
    P'^{(g)}(x,y) = \frac{P^{(g)}(x,y)}{max(|P^{(g)}(x,y)|)} 
\end{equation}
before displaying it on the SLM. 
However, this normalization leads to mode-dependent losses, since the modes are no longer normalized according to $\int P P^* dxdy = 1 $. This issue can be corrected either by postprocessing the measured counts or by pre-multiplying the modes by suitable scaling factors before displaying them on the SLM such that $\int P_j P_j^* dxdy $ is equal for the state $\ket{\nu}$ and all the gauging states $\ket{\phi_j}$ forming the measurement basis.

\begin{figure}[h!]
    \centering
    \includegraphics[width=0.7\linewidth]{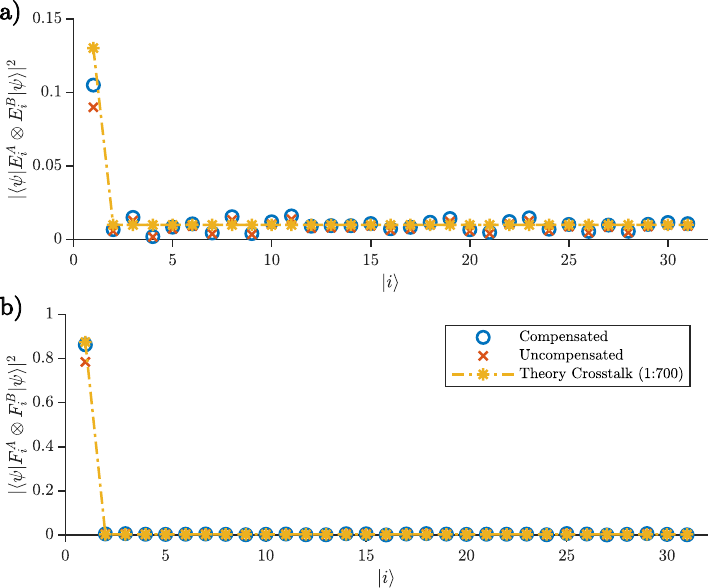}
    \caption{Normalization of coincidence counts using calibrated counts when Alice and Bob are projecting on the same state. The input state and measuring basis correspond to the average fidelity model with $\varepsilon = 0.007$. a) Projection of the optimal faking state $\ket{\psi}$ on the computational basis $E^A_i\otimes E^B_i$. b) Projection of $\ket{\psi}$ in the Fourier basis.} 
    \label{fig:Bin_coinmeas}
\end{figure}

In the case of coincidence counting, one can follow the same procedure and arrive at
\begin{equation}\label{eq:Bin_Coinmodlosses}
    \frac{C_{\nu \nu \, i j}}{C_{ii\,jj}} = \frac{\eta_\nu^2 \,\text{Tr}\left( \tilde{E}_i \ketbra{\nu}{\nu}\otimes\tilde{E}_j \ketbra{\nu}{\nu}\right)}{ \eta_i \eta_i},
\end{equation}
where $C_{ii\,jj}$ is the coincidence count rate for a generated state $\ket{\phi_i}\otimes\ket{\phi_i}$ projected on $E_j\otimes E_j$ in the measurement setup. 
Fig.~\ref{fig:Bin_coinmeas} shows the measured detection probabilities of the optimal faking state $\ket{\psi}$ obtained from the coincidence counts $\frac{C_{\nu \nu \, j j}}{C_{jj\,jj}}$ with and without compensation for efficiencies.

The advantage of employing calibrated probabilities is that we no longer need to measure all $2 d^2$ possible outcomes of a two-photon $d$-dimensional state in order to evaluate the entanglement witness in Eq.~(\ref{eq:SN}), as required by the standard normalization of count rates in multi-outcome measurements. Instead, we only need $2d$ calibration measurements for the Fourier and computational bases, along with $2d$ measurements of the target state to calculate the entanglement witness.
It is important to reiterate that detection efficiencies do not pose a problem for the multi-outcome measurements, since the state efficiency $\eta_\nu$ acts as a global loss factor that is absorbed into the normalization constant $\Phi_T$.

\subsection{Experimental results for the worst-case attacks}
\begin{figure}[ht]
    \centering
    \includegraphics[width=1\linewidth]{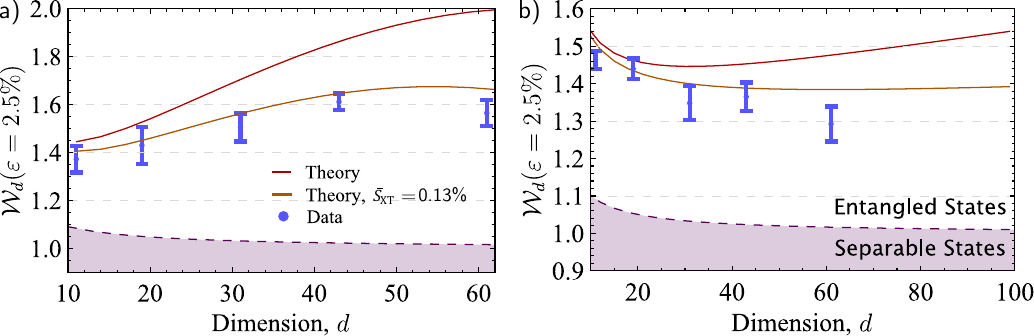}
    \caption{Comparison between the experimental data and theoretical predictions for a) the worst-case fidelity models, and b) the worst-case fidelity with an individual projector error model. The red curve indicates the ideal theoretical model, whereas the orange curve shows the model including an average per-mode crosstalk $\bar{S}_{XT}= 0.13\%$.}
    \label{fig:SingleandBinCurves}
\end{figure}

The average fidelity model for measurement errors introduced above concentrates a large amount of the error into a single projector. 
This allows a skeptical device-user to easily spot an attack by carefully inspecting the crosstalk measurements shown in Fig.~\ref{fig:Channel_average}. 
As indicated in Sec.~\ref{sec:avfid_classicalrandom}, this asymmetry can be overcome by leveraging ordinary classical randomness between Alice's and Bob's devices, which only needs to be supplied before starting the experiment. 
Nevertheless, whether classical randomness is employed or not, there is more than one ways for device-users to benchmark the fidelity error of the measurements.
The choice among these can be motivated for example by the chosen experimental implementation. 
As mentioned before and \ref{sec:BinProcess}, we have additionally designed attacks for scenarios where the measurement fidelity is quantified for i) the worst-case fidelity of a multi-outcome measurement (see Sec.~\ref{sec:worst_case}), and ii) the worst-case fidelity when a multi-outcome measurement is implemented through a sequence of single projections --- which is a particularly common practice in photonics, see Sec.~\ref{sec:Bin_worst_case}.
From the worst-case nature of these quantifiers follows that for every outcome $l$, $\langle O_l|\tilde{O}_l|O_l\rangle \geq 1-\varepsilon$ holds. 
This is a more stringent benchmark than the average fidelity discussed before. Furthermore, this fidelity constrain removes any asymmetry from the attack, thereby overcoming the weakness of the average fidelity attack. 

For case i), we show the entanglement witness for different dimensions with $\varepsilon=2.5\%$  in Fig.~\ref{fig:SingleandBinCurves}\,a), which results in falsely certifying an entanglement dimensionality of up to 32 for $d=61$. 
For case ii), the measurement restricted to only individual projections instead of true multi-outcome measurement. 
This means that by design the model perform worse for some intermediate-range dimensions. However, the model become much more powerful in very high dimensions because they can additionally leverage the separate implementation of the individual projectors to further undermine the entanglement certification.
The experimental results up to $d=61$ are shown in Fig.~\ref{fig:SingleandBinCurves}\,b) for $\varepsilon = 2.5\%$ resulting in faking an entanglement dimensionality of up to 16.
The intrinsic crosstalk in our system prevents us from demonstrating the extra leverage in attack ii) in higher dimensions in comparison to the other models.

We show that both attack models lead to qualitatively similar conclusions as the average fidelity model, particularly regarding both attack performance with dimension $d$ and perturbation $\varepsilon$. 

\subsection{Extra measurements: Entanglement-separable gap}

The entanglement to separable gap $\Delta$ was defined in section~\ref{sec:enttosep}. The gap in Eq.~(\ref{eq:entsepgap}) can be directly evaluated from the value of the entanglement witness $\mathcal{W}_{\text{prod}}^{(d)}(\varepsilon)$ for the optimal faking state at a given perturbation $\varepsilon$. 
Figs.~\ref{fig:EntGap_average}, \ref{fig:EntGap_independent}, and \ref{fig:EntGap_binarized} display the observed dependence of the gap $\Delta$ on the perturbation $\varepsilon$ and the dimension $d$ for the average fidelity model, the worst-case attack, and the worst-case attack with individual projectors, respectively.

\begin{figure}[h!]
    \centering
    \includegraphics[width=0.7\linewidth]{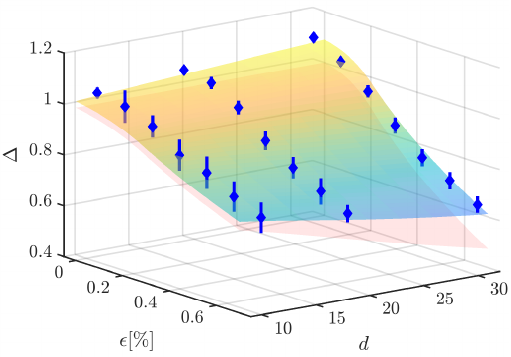}
    \caption{Entanglement-to-separable state gap $\Delta$ for the average fidelity model. Blue dots represent measurements, the yellow to blue surface is the theoretical prediction assuming an average per-mode crosstalk $\bar{S}_{XT}= 0.115\%$; and the red surface is the ideal theory.}
    \label{fig:EntGap_average}
\end{figure}

\begin{figure}[h!]
    \centering
    \includegraphics[width=0.7\linewidth]{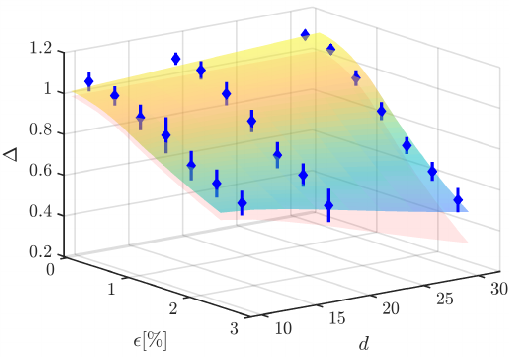}
    \caption{Entanglement-to-separable state gap $\Delta$ for the worst-case fidelity model. Blue dots represent measurements, the yellow to blue surface is the theoretical prediction assuming an  average per-mode crosstalk $\bar{S}_{XT}= 0.13\%$; and the red surface is the ideal theory.}
    \label{fig:EntGap_independent}
\end{figure}

\begin{figure}[h!]
    \centering
    \includegraphics[width=0.7\linewidth]{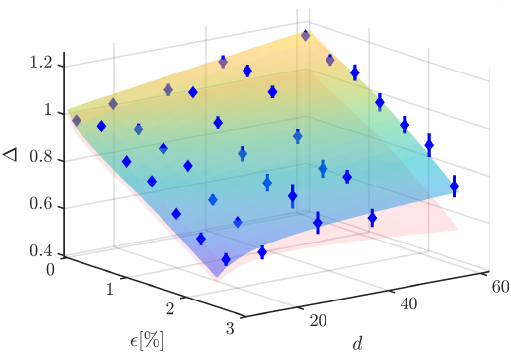}
    \caption{Entanglement-to-separable state gap $\Delta$ for the worst-case fidelity model with individual projections. Blue dots represent measurements, the yellow to blue surface is the theoretical prediction assuming an average per-mode crosstalk $\bar{S}_{XT}= 0.13\%$; and the red surface is the ideal theory.}
    \label{fig:EntGap_binarized}
\end{figure}

}

\newpage
